\documentclass[pdflatex,sn-mathphys-ay]{sn-jnl}

\usepackage{graphicx}%
\usepackage{multirow}%
\usepackage{amsmath,amssymb,amsfonts}%
\usepackage{amsthm}%
\usepackage{mathrsfs}%
\usepackage[title]{appendix}%
\usepackage{xcolor}%
\usepackage{textcomp}%
\usepackage{manyfoot}%
\usepackage{booktabs}%
\usepackage{algorithm}%
\usepackage{algorithmicx}%
\usepackage{algpseudocode}%
\usepackage{listings}%

\theoremstyle{thmstyleone}%
\newtheorem{theorem}{Theorem}
\newtheorem{proposition}[theorem]{Proposition}%

\theoremstyle{thmstyletwo}%
\newtheorem{remark}{Remark}%

\theoremstyle{thmstylethree}%

\hypersetup{
  hypertexnames=false,
  pdftitle={Decoupled Probabilistic Forecasting and Arbitrage-Aware Refinement of Implied Volatility Surfaces},
  pdfauthor={Lifeng Hao and Shaolin Ji},
  pdfsubject={Implied volatility surface forecasting},
  pdfkeywords={Implied volatility surface, probabilistic forecasting,
    static no-arbitrage, conditional diffusion, surface-aware attention}
}

\begin{document}

\title[Arbitrage-Aware IVS Forecasting]{Decoupled Probabilistic Forecasting and Arbitrage-Aware Refinement of Implied Volatility Surfaces}

\author*[1]{\fnm{Lifeng} \sur{Hao}}\email{201911961@mail.sdu.edu.cn}
\equalcont{These authors contributed equally to this work.}

\author[1]{\fnm{Shaolin} \sur{Ji}}\email{jsl@sdu.edu.cn}
\equalcont{These authors contributed equally to this work.}

\affil*[1]{\orgdiv{Zhongtai Securities Institute for Financial Studies}, \orgname{Shandong University}, \orgaddress{\city{Jinan}, \postcode{250100}, \state{Shandong}, \country{China}}}

\abstract{
Implied volatility surface forecasting is essential for option valuation, hedging, and risk management, but 
remains difficult because future surfaces are stochastic while pricing inputs must satisfy static no-arbitrage 
shape restrictions. We propose a decoupled generative refinement framework for IVS forecasting as an operational 
risk surface modeling problem. The first stage uses a conditional diffusion model to learn the conditional distribution 
of future surfaces. The generated ensemble captures predictive distributional variation, and its median provides a 
robust representative surface for subsequent refinement. The second stage introduces a Surface Aware Attention 
Module (SAAM), a cross sectional refinement operator that improves fit to market observations and static no-arbitrage 
diagnostics for the representative surface. This design separates distribution learning from surface refinement, 
allowing the diffusion model to capture stochastic market dynamics while SAAM controls static no-arbitrage residual 
violations on the final surface. We evaluate the framework on CSI 300 index options from June 2020 to September 2024 
under daily and minute level forecasting protocols. The diffusion stage improves forecasting accuracy and produces 
predictive intervals that vary across moneyness, maturity, and sampling frequency. The refinement stage improves 
fitting accuracy against market observations and reduces measured static no-arbitrage residual violations, with 
stronger gains at the minute level. Attention diagnostics suggest that SAAM performs adaptive cross sectional 
refinement rather than fixed local smoothing.
}



\maketitle

\section{Introduction}
\label{sec:intro}
The reliable modeling of option implied volatility surfaces (IVS) is a central problem in quantitative finance and risk management. 
IVS forecasts are key inputs for derivative valuation, hedging, portfolio risk measurement, and volatility trading. Unlike conventional 
financial time series, an IVS evolves over a moneyness and maturity domain and must satisfy static no arbitrage 
restrictions \citep{dupire1994pricing,heston1993closed,Gatheral2006}. Accurate modeling therefore requires both stochastic predictive 
flexibility and consistency with static no arbitrage restrictions. We represent the IVS observed at time $t$ as a two dimensional surface $\Sigma_t$
over moneyness $m$ and time to maturity $\tau$.

\begin{figure}[htbp]
\centering
\includegraphics[width=0.55\linewidth]{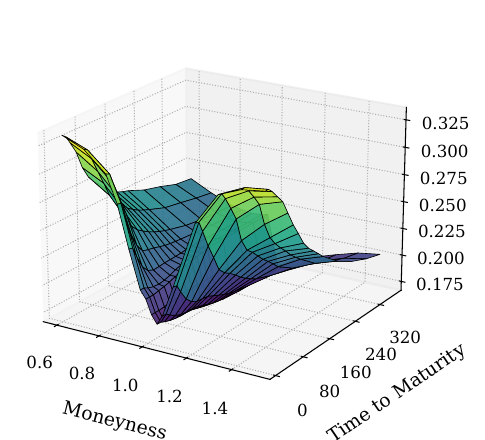}
\caption{Typical implied volatility surface (IVS) for CSI 300 options. The z-axis denotes the volatility level.}
\label{fig:intro_average_ivs}
\end{figure}

As illustrated in Figure~\ref{fig:intro_average_ivs}, empirical IV surfaces exhibit nonlinear structures such as term structures, volatility skews, and smiles \citep{FRANKS1991,HEYNEN1994,Dumas1998,Azzone2025,zaugg2025}. Their dynamics are also affected by liquidity heterogeneity, asymmetric tail expectations, and latent market shocks \citep{Cont2002a}. Consequently, future IVS states are better viewed as conditional distributions of possible market scenarios rather than as single deterministic trajectories.

A key challenge is to reconcile stochastic scenario generation with static no-arbitrage requirements. Flexible data-driven models can capture complex predictive 
dynamics, but may generate surfaces with static-arbitrage residual violations. Conversely, imposing geometric penalties within a forecasting objective is 
nontrivial because empirical IVS observations may contain transient violations caused by uneven contract coverage, interpolation effects, bid-ask frictions, and 
microstructure noise. This creates a trade-off between empirical fidelity and static no-arbitrage residual control, especially in sparse, tail-sensitive, and 
high-frequency settings.

Motivated by this optimization tension, we propose a decoupled probabilistic framework for IVS scenario generation and surface refinement with static no-arbitrage 
restrictions. The first stage uses a conditional diffusion model \citep{Ho2020,song2021scorebased,tashiro2021csdi} to learn the conditional 
distribution of future IVS states. It transforms IVS forecasting from deterministic surface extrapolation into probabilistic 
scenario generation. The generated ensemble is retained for distributional evaluation, and its median provides a robust representative forecast surface for Stage II refinement.

The second stage introduces the Surface Aware Attention Module (SAAM), a coordinate based refinement operator. SAAM maps the representative forecast 
surface to arbitrary target coordinates, allowing supervision from scattered market observations and residual diagnostics. 
By aggregating information from the entire representative surface, SAAM exploits the cross sectional coupling of smiles and term structures rather than relying on 
isolated pointwise correction. The refined surface is designed to improve fit to market observations and static no-arbitrage residual control while preserving the 
Stage I predictive signal.

The empirical evaluation follows this two stage design. Stage I is assessed by forecasting accuracy and distributional forecast quality for future IVS states. 
Stage II is evaluated as a surface level refinement task, using fit to market observations and static no-arbitrage diagnostics for the refined representative 
surface. From an operational perspective, this design separates the generation of future risk surface scenarios from the construction of a refined representative 
surface used for option valuation, risk measurement, and static no arbitrage assessment.
The main contributions of this work are as follows:
\begin{enumerate}
    \item \textbf{A diffusion based probabilistic formulation for IVS forecasting.}
    We introduce conditional diffusion into IVS forecasting to model the conditional distribution of future implied volatility surfaces. 
    The model generates predictive surface ensembles for distributional evaluation, and their median provides a robust representative forecast surface for subsequent refinement.

    \item \textbf{A surface aware attention refinement operator.}
    We introduce SAAM as a coordinate based refinement module that maps a representative forecast surface to arbitrary target coordinates, 
    allowing fitting against scattered market observations and evaluation on synthetic diagnostic grids. Its attention mechanism enables nonlocal information 
    aggregation across the moneyness and maturity domain and supports residual based regularization for static no-arbitrage diagnostics.
    
    \item \textbf{Empirical evidence from CSI 300 index options.}
    Using daily and minute level CSI 300 index option data, we show that the diffusion stage generates informative predictive surface ensembles, 
    while the SAAM stage improves fit to market observations and  static no-arbitrage residual control. The learned attention patterns further 
    provide diagnostic evidence on cross sectional information allocation across tail sensitive and maturity dependent regions.
    
\end{enumerate}

\section{Related Work}
\label{sec:related_work}

\subsection{Classical Parameterizations and Stochastic Dynamic Modeling}
Implied volatility surface (IVS) modeling has traditionally relied on static parametric specifications and continuous time stochastic processes. 
Classical parameterizations, such as the Stochastic Volatility Inspired (SVI) family \citep{Gatheral2014}, provide parsimonious 
representations of IVS and can be made consistent with static no-arbitrage conditions under suitable parameter restrictions. 
However, they are primarily designed for cross sectional fitting or interpolation and do not by themselves provide an endogenous mechanism for 
forecasting future surface dynamics. To capture temporal evolution, market models employ factor structures or stochastic differential 
equations \citep{dewynne1999market,Durrleman2010,Audrino2010,ait2021implied,Cont2023}. While theoretically rigorous and closely connected to 
arbitrage-free pricing theory, these approaches often rely on restrictive assumptions about the driving factors, volatility dynamics, or functional 
form of surface evolution. Their implementation can therefore become challenging when observed contracts are sparse and unevenly distributed across 
the moneyness and maturity domain, and when market observations are affected by regime dependent fluctuations.

\subsection{Deep IVS Forecasting and Penalty Based Methods}

The limitations of classical parametric and stochastic dynamic models have motivated the use of deep learning methods for IVS forecasting. 
Convolutional neural networks and recurrent architectures, such as LSTMs, provide data driven forecasting benchmarks by learning nonlinear 
temporal dependencies from historical IVS \citep{Medvedev2022,Zhang2022,Almeida2023,Olsen2025,Shao2025}. However, when trained 
mainly with mean squared error objectives, these models tend to produce conditional mean forecasts, which may smooth future surface movements and 
underrepresent scenario diversity during regime shifts or tail risk repricing. In their standard form, they also do not explicitly model the conditional 
distribution of future IVS states, limiting their ability to represent market stochasticity in scenario based risk analysis.

To improve financial consistency, recent studies often embed soft penalties for static no-arbitrage conditions into neural network 
losses \citep{Ackerer2020,Zheng2021,Zhang2022}. Such penalty based formulations are practical and empirically useful, and the refinement 
stage developed in this paper also adopts residual based regularization for static no-arbitrage restrictions. The main limitation is therefore 
not the use of soft penalties per se, but the frequent reliance on pointwise mappings such as multi layer perceptrons. In these coordinate to 
volatility architectures, the output at each target coordinate is generated mainly from local input features and is only indirectly coupled with 
other surface regions through shared parameters and global loss terms. This can be restrictive for IVS refinement, since option smiles, term structures, 
and static no-arbitrage conditions are intrinsically cross sectional.

The closest methodological benchmark is the two step IVS prediction framework with static no-arbitrage penalties proposed by \citet{Zhang2022}, 
which separates temporal prediction from surface construction and evaluates residual penalties on synthetic grids. We build on this decomposition 
principle and residual collocation logic, but use conditional diffusion to generate predictive IVS scenarios rather than a single forecast surface. 
At the surface refinement stage, we use SAAM instead of a pointwise MLP mapping. This allows each target coordinate to aggregate nonlocal information 
from the representative forecast surface, thereby exploiting the cross sectional coupling of smiles and term structures during refinement.

\subsection{Generative Scenario Modeling}

Generative models provide a natural route beyond single surface forecasts, since financial risk analysis often requires multiple plausible future market states 
rather than a single conditional mean forecast. In option markets, VolGAN \citep{Vuletić2025} provides an important example by simulating joint scenarios of the underlying 
asset and the implied volatility surface. VolGAN incorporates smoothness regularization into the generator objective, while static no-arbitrage considerations are 
handled mainly through ex-post scenario reweighting. This reweighting mechanism changes the probability weights over generated 
scenarios rather than directly modifying a given surface.

The broader generative modeling literature provides further motivation for a diffusion based approach. Although GAN based models are effective for scenario simulation, 
adversarial training can be sensitive to instability and mode collapse \citep{Thanh2020,Jabbar2021}. Diffusion models offer a non adversarial alternative by learning 
to reverse a stochastic noising process, with score based formulations providing a closely related continuous time perspective \citep{Ho2020,song2021scorebased}. 
These models have demonstrated strong sample quality and distributional coverage in high-dimensional generation \citep{Dhariwal2021}. Initially developed for high 
dimensional  data generation \citep{yangsong2023,harvery2022video,Ho2022video,Avrahami2022multimode}, diffusion based models have recently been adapted to 
conditional time series and spatiotemporal forecasting \citep{tashiro2021csdi,timeseries}.

This study addresses a complementary layer of the IVS forecasting problem. We use conditional diffusion to generate predictive ensembles of future IVS states, 
and retain these ensembles for distributional evaluation. Instead of reweighting generated scenarios, we extract a representative 
forecast surface and refine it directly through SAAM. The refinement stage uses attention based nonlocal information aggregation and residual regularization for static 
no-arbitrage diagnostics, allowing distributional scenario generation and representative surface refinement to be handled separately.

\section{Problem Formulation and No-Arbitrage Constraints}
\label{sec:problem_formulation}

This section provides essential background on the implied volatility surface (IVS).

\subsection{Problem Formulation and State Representation}

Implied volatility is defined implicitly via an option pricing formula. For a European call option with strike $K$ and time-to-maturity $\tau=T-t$, 
let $F_t(\tau)$ and $D_t(\tau)$ denote the corresponding forward index level and discount factor. Under the Black--76 convention, the pricing function is
\begin{equation}
C_{\mathrm{B76}}(F_t(\tau),K,\tau,\sigma)
=
D_t(\tau)
\left[
F_t(\tau) N(d_1)-K N(d_2)
\right],
\end{equation}
where
\begin{align}
d_1
&=
\frac{
\ln(F_t(\tau)/K)+\frac{1}{2}\sigma^2\tau
}{
\sigma\sqrt{\tau}
},
\qquad
d_2
=
d_1-\sigma\sqrt{\tau},
\end{align}
and $N(\cdot)$ is the standard normal cumulative distribution function. 

Given an observed market price $C_t^{\mathrm{mkt}}(K,T)$, the implied volatility $\sigma_t(K,T)$ is defined as:
\begin{equation}
C_{\mathrm{B76}}
\bigl(
F_t(\tau),K,\tau,\sigma_t(K,T)
\bigr)
=
C_t^{\mathrm{mkt}}(K,T).
\end{equation}

For surface modeling, we reparameterize implied volatility using log-forward moneyness $m=\ln(K/F_t(\tau))$ and maturity $\tau$. The standardized implied 
volatility surface at time $t$ is thus represented as
\begin{equation}
\Sigma_t(m,\tau)
:=
\sigma_t(F_t(\tau)e^m,t+\tau).
\end{equation}
For a fixed coordinate $(m,\tau)$, $\Sigma_t(m,\tau)$ evolves over time as a stochastic process; for a fixed date $t$, $\Sigma_t(\cdot,\cdot)$ 
describes the prevailing cross-sectional option risk state. 

\subsection{Static No-Arbitrage Conditions}

Conditions that ensure an IVS is free of static arbitrage have been well investigated in the literature \citep{Lee2004, Roper2010, Gatheral2014}. 
Following this theoretical framework, we state these conditions in terms of the total implied variance:
\begin{equation}
\phi(m,\tau) = \Sigma(m,\tau)^2\tau.
\end{equation}

\begin{proposition}[Static arbitrage-free conditions]
\label{prop:geometric_constraints}
Consider an IVS $\Sigma(m,\tau)$ and its total implied variance $\phi(m,\tau)=\Sigma(m,\tau)^2\tau$. Suppose that the following conditions are satisfied:
\begin{enumerate}
\item \textit{(Positivity)} $\phi(m,\tau) > 0$ for every $(m, \tau)$.
\item \textit{(Zero-Maturity Limit)} $\lim_{\tau \downarrow 0}\phi(m,\tau)=0$ for every $m \in \mathbb{R}$.
\item \textit{(Differentiability)} For every $\tau > 0$, $m \mapsto \phi(m,\tau)$ is twice differentiable, and for every $m \in \mathbb{R}$, $\tau \mapsto \phi(m,\tau)$ is differentiable.
\item \textit{(Monotonicity)} For every $m \in \mathbb{R}$, $\tau \mapsto \phi(m,\tau)$ is non-decreasing, yielding the calendar spread condition:
\begin{equation}
\ell_{\mathrm{cal}}(m,\tau) = \partial_\tau \phi(m,\tau) \geq 0.
\end{equation}
\item \textit{(Durrleman's Condition)} For every $(m,\tau)$, the butterfly spread condition holds:
\begin{equation}
\ell_{\mathrm{but}}(m,\tau) = \left( 1-\frac{m\partial_m\phi}{2\phi} \right)^2 - \frac{(\partial_m\phi)^2}{4} \left( \frac{1}{\phi}+\frac{1}{4} \right) + \frac{\partial_{mm}^2\phi}{2} \geq 0.
\end{equation}
\item \textit{(Tail Asymptotics)} For every $\tau > 0$, $\phi(m,\tau) = \mathcal{O}(|m|)$ as $|m| \to \infty$.
\end{enumerate}
Then, $\Sigma(m,\tau)$ is free of static arbitrage.
\end{proposition}

Condition 4 ensures the absence of calendar spread arbitrage, while Condition 5 guarantees the absence of butterfly arbitrage and the admissibility of 
the cross-sectional risk neutral density. These two quantities, $\ell_{\mathrm{cal}}$ and $\ell_{\mathrm{but}}$, thus provide the central  
diagnostics for  arbitrage violations in our empirical assessment.

\subsection{Continuous Surface Reconstruction from Scattered Market Observations}

In listed option markets, the available strikes and maturities vary across dates as contracts are introduced, expire, or become inactive. 
Consequently, at time $t$, implied volatilities are observed only at an irregular and time-varying set of contract locations:

\begin{equation}
\Gamma_t
=
\left\{
\left(m_i,\tau_i,\Sigma_t(m_i,\tau_i)\right)
\mid
i=1,\ldots,n_t
\right\}.
\end{equation}
Since both the number of observations and their locations vary over time, a continuous surface representation is required before constructing a comparable IVS state across dates.

We reconstruct the surface from scattered observations using the Nadaraya--Watson (NW) kernel estimator~\citep{Hardle1990}. For any target 
coordinate $(m,\tau)$, the reconstructed implied volatility surface is defined as
\begin{equation}
\label{eq:NW_estimator}
\hat{\Sigma}_t(m,\tau)
=
\frac{
\sum_{i=1}^{n_t}
\Sigma_t(m_i,\tau_i)
g_{h_1,h_2}(m-m_i,\tau-\tau_i)
}{
\sum_{i=1}^{n_t}
g_{h_1,h_2}(m-m_i,\tau-\tau_i)
},
\end{equation}
where $g_{h_1,h_2}(\cdot,\cdot)$ is a bivariate Gaussian kernel with bandwidths $h_1$ and $h_2$. The bandwidths determine the degree of smoothing in the moneyness 
and maturity directions and are selected date by date using least squares cross validation. The resulting reconstructed field $\hat{\Sigma}_t(m,\tau)$ provides a 
continuous market state representation.

\subsection{Structured Spatial Discretization}
\label{sec:data_grid}

To map the continuous reconstructed surface $\hat{\Sigma}_t$ into a standardized finite dimensional input for generative modeling, we discretize the surface onto a 
fixed moneyness--maturity grid $\mathcal{D}_0$. We restrict our domain to liquid regions by retaining out-of-the-money options within $m \in [\log(0.6), \log(1.5)]$ 
and $\tau \geq 10/365$. The grid is defined as $\mathcal{D}_0 = \{ (m, \tau) \mid m \in \mathcal{M}_0, \tau \in \mathcal{T}_0 \}$, where:
\begin{align*}
\mathcal{M}_0 &= \{ \log(x) \mid x \in \{0.6, 0.7, 0.8, 0.85, 0.9, 0.95, 0.975, 1, 1.025, 1.05, 1.1, 1.2, 1.3, 1.4, 1.5\} \},\\
\mathcal{T}_0 &= \{ i/365 \mid i \in \{10, 20, 30, 45, 60, 75, 91, 106, 122, 137, 152, 167, 182, 227, 273, 319, 365\} \}.
\end{align*}
Evaluating $\hat{\Sigma}_t$ on these $15 \times 17$ spatial nodes yields the fixed grid state $\bar{\Sigma}_t \in \mathbb{R}^{15 \times 17}_{+}$. This tensor 
sequence $\{\bar{\Sigma}_t\}$ serves as the input to the generative model, while the original observations $\Gamma_t$ are retained separately for Stage-II data 
fidelity supervision. Building on this standardized representation and the diagnostics in Proposition~\ref{prop:geometric_constraints}, Section~\ref{sec:methodology} 
introduces our decoupled generation and refinement framework.

\section{Methodology}
\label{sec:methodology}

\begin{figure*}[htbp]
\centering
\includegraphics[width=0.9\textwidth]{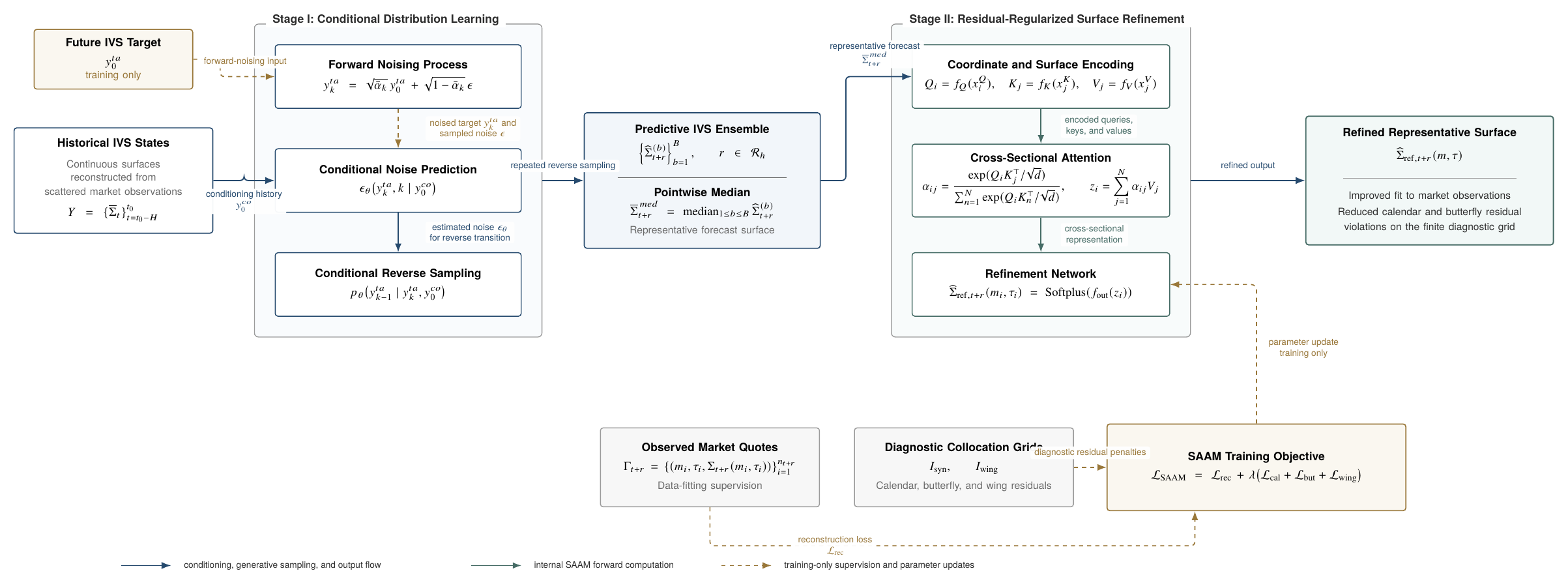}
\caption{\textbf{Overall architecture of the proposed decoupled IVS framework.}}
\label{fig:architecture}
\end{figure*}

This section presents the proposed decoupled framework for conditional IVS
scenario generation and representative surface refinement. As shown in
Figure \ref{fig:architecture}, Stage I learns the conditional distribution of future IVS states
and generates an ensemble of possible risk surface scenarios. The full
ensemble is retained for distributional evaluation, while its pointwise
median provides a robust representative forecast for Stage II.
Stage II refines this representative surface against scattered market
observations. Its training objective combines data fitting with residual
regularization for the calendar, butterfly, and wing diagnostics evaluated
on synthetic collocation grids. This decomposition separates conditional
distribution learning from surface refinement, allowing the diffusion model
to capture stochastic market dynamics while SAAM improves fit to market
observations and reduces finite-grid static no-arbitrage residual violations
for the final representative surface.

\subsection{Stage I: Conditional Scenario Generation via Diffusion}
\label{sec:stage1_diffusion}
The $\mathcal{D}_0$ is the fixed moneyness--maturity grid with $N=|\mathcal{D}_0|=255$. For notational convenience, we vectorize each fixed grid IVS state as
$y_t = \operatorname{vec}(\bar{\Sigma}_t) \in \mathbb{R}_{+}^{N}$.
Given a historical window of length $H$ and a forecast path length $h$, define the horizon index set
$\mathcal{R}_h=\{1,\ldots,h\}$.
We construct the conditioning block and the future target block as
\begin{align}
y_0^{\mathrm{co}} &= \left[ y_{t-H+1},\ldots,y_t \right] \in \mathbb{R}_{+}^{N\times H}, \label{eq:conditioning_window} \\
y_0^{\mathrm{ta}} &= \left[ y_{t+r} \right]_{r\in\mathcal{R}_h} \in \mathbb{R}_{+}^{N\times h}. \label{eq:target_surface}
\end{align}
When $h=1$, the target reduces to the next fixed grid IVS state. When $h>1$, the target represents the future IVS path over the next $h$ horizons. 
Stage I approximates the conditional predictive distribution
$p_{\mathrm{data}}(y_0^{\mathrm{ta}}\mid y_0^{\mathrm{co}})$
by a parameterized generative model
$p_{\theta}(y_0^{\mathrm{ta}}\mid y_0^{\mathrm{co}})$.
We use $k=0,\ldots,K$ for the diffusion index, which is distinct from the time index $t$ and the forecast horizon index $r$.

\subsubsection{Forward diffusion process}

The forward process perturbs the future target block into Gaussian noise through the  Markov chain
\begin{equation}
q(y_{1:K}^{\mathrm{ta}}\mid y_0^{\mathrm{ta}}) = \prod_{k=1}^{K} q(y_k^{\mathrm{ta}}\mid y_{k-1}^{\mathrm{ta}}),
\label{eq:forward_process_chain}
\end{equation}
where
\begin{equation}
q(y_k^{\mathrm{ta}}\mid y_{k-1}^{\mathrm{ta}}) = \mathcal{N} \left( y_k^{\mathrm{ta}}; \sqrt{1-\beta_k}y_{k-1}^{\mathrm{ta}}, \beta_k I_{D_h} \right).
\label{eq:forward_process_step}
\end{equation}
Here $\beta_k\in(0,1)$ is a pre-specified variance schedule and $I_{D_h}$ is the $D_h$-dimensional identity matrix. Let $\alpha_k=1-\beta_k$ and $\bar{\alpha}_k=\prod_{s=1}^{k}\alpha_s$. Then
\begin{equation}
y_k^{\mathrm{ta}} = \sqrt{\bar{\alpha}_k}y_0^{\mathrm{ta}} + \sqrt{1-\bar{\alpha}_k}\boldsymbol{\epsilon}, \qquad \boldsymbol{\epsilon}\sim\mathcal{N}(0,I_{D_h}).
\label{eq:forward_sample}
\end{equation}

\subsubsection{Conditional reverse denoising process}

The reverse process recovers the future target block from the noisy state while conditioning on the historical IVS trajectory:
\begin{equation}
p_{\theta}(y_{0:K}^{\mathrm{ta}}\mid y_0^{\mathrm{co}}) = p(y_K^{\mathrm{ta}}) \prod_{k=1}^{K} p_{\theta}(y_{k-1}^{\mathrm{ta}}\mid y_k^{\mathrm{ta}},y_0^{\mathrm{co}}),
\label{eq:reverse_process_chain}
\end{equation}
where $p(y_K^{\mathrm{ta}})=\mathcal{N}(0,I_{D_h})$ and
\begin{equation}
p_{\theta}(y_{k-1}^{\mathrm{ta}}\mid y_k^{\mathrm{ta}},y_0^{\mathrm{co}}) = \mathcal{N} \left( y_{k-1}^{\mathrm{ta}}; \boldsymbol{\mu}_{\theta}(y_k^{\mathrm{ta}},k\mid y_0^{\mathrm{co}}), \tilde{\beta}_k I_{D_h} \right).
\label{eq:reverse_process_step}
\end{equation}
The conditioning history $y_0^{\mathrm{co}}$ is kept noise free, so that the model learns the conditional transition from historical surfaces to future surface scenarios without corrupting observed market-state information.
Using the standard noise prediction parameterization, the reverse mean is
\begin{equation}
\boldsymbol{\mu}_{\theta}(y_k^{\mathrm{ta}},k\mid y_0^{\mathrm{co}}) = \frac{1}{\sqrt{\alpha_k}} \left( y_k^{\mathrm{ta}} - \frac{\beta_k}{\sqrt{1-\bar{\alpha}_k}} \boldsymbol{\epsilon}_{\theta}(y_k^{\mathrm{ta}},k\mid y_0^{\mathrm{co}}) \right),
\label{eq:reverse_mean}
\end{equation}
with posterior variance
\begin{equation}
\tilde{\beta}_k = \begin{cases}
\dfrac{1-\bar{\alpha}_{k-1}}{1-\bar{\alpha}_k}\beta_k, & k>1, \\
\beta_1, & k=1.
\end{cases}
\label{eq:reverse_var}
\end{equation}
The diffusion model is trained by minimizing
\begin{equation}
\mathcal{L}_{\mathrm{diff}} = \mathbb{E}_{y_0^{\mathrm{ta}},y_0^{\mathrm{co}},\boldsymbol{\epsilon},k} \left[ \left\| \boldsymbol{\epsilon}_{\theta}(y_k^{\mathrm{ta}},k\mid y_0^{\mathrm{co}}) - \boldsymbol{\epsilon} \right\|_2^2 \right].
\label{eq:diffusion_loss}
\end{equation}

After training, repeated reverse sampling gives a predictive ensemble of future IVS paths. For each sampled path and each horizon $r\in\mathcal{R}_h$, the corresponding vector is reshaped onto $\mathcal{D}_0$, yielding
\begin{equation}
\left\{ \widehat{\Sigma}_{t+r}^{(b)} \right\}_{b=1}^{B}, \qquad r\in\mathcal{R}_h .
\label{eq:predictive_ensemble}
\end{equation}
To interface the generative stage with baselines and the subsequent refinement stage, we extract a representative forecast surface at each horizon by taking the pointwise ensemble median:
\begin{equation}
\widehat{\Sigma}_{t+r}^{\mathrm{med}}(m,\tau) = \operatorname{median}_{1\leq b\leq B} \widehat{\Sigma}_{t+r}^{(b)}(m,\tau), \qquad r\in\mathcal{R}_h .
\label{eq:ensemble_median}
\end{equation}
Thus, the diffusion model is not reduced to a single IVS forecast: its ensemble output supports distributional evaluation, while the median provides a stable representative surface for Stage II refinement.

\subsection{Stage II: Surface Refinement with Static No-Arbitrage Restrictions}
\label{sec:SAAM}
The diffusion stage generates predictive surface scenarios that need not satisfy the static no-arbitrage restrictions in Proposition~\ref{prop:geometric_constraints}. 
We introduce the Surface Aware Attention Module (SAAM) as a coordinate based refinement operator. In our empirical implementation, SAAM is applied horizon by horizon to 
the pointwise ensemble median $\widehat{\Sigma}_{t+r}^{\mathrm{med}}$, yielding a refined representative surface for  comparison and static no-arbitrage  evaluation. 

\begin{figure}[htbp]
\centering
\includegraphics[width=0.5\textwidth]{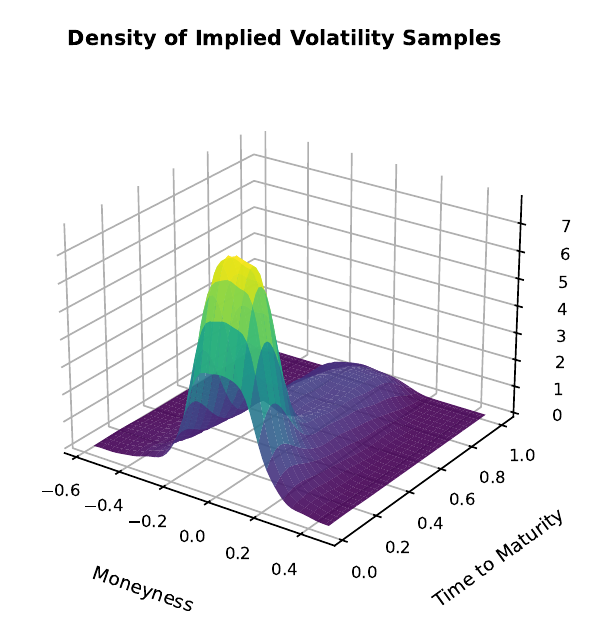}
\caption{Empirical density of observed option contracts over the moneyness and maturity domain for CSI 300 index options.}
\label{fig:csi300_implied_volatility_density}
\end{figure}

The design of SAAM is motivated by the uneven spatial coverage of observed option contracts. As shown in Figure~\ref{fig:csi300_implied_volatility_density}, contracts cluster 
near-the-money and at-short-maturities, while wing and long maturity regions are more sparsely observed. Since calendar and butterfly restrictions involve cross-sectional 
relations across maturity and moneyness, a refinement rule based only on local observations may be unreliable in sparse regions. 
SAAM instead uses the fixed grid representative surface as a global reference and allows each target coordinate to aggregate information from all grid nodes. 
The contract density pattern motivates this global refinement design, but is not used as an explicit weighting scheme.
For a given horizon $r\in\mathcal{R}_h$, let

\begin{equation}
\widetilde{\Sigma}_{t+r}(m_j,\tau_j) = \widehat{\Sigma}_{t+r}^{\mathrm{med}}(m_j,\tau_j), \qquad (m_j,\tau_j)\in\mathcal{D}_0,
\label{eq:representative_surface}
\end{equation}
denote the unrefined representative forecast surface. SAAM uses the fixed grid nodes of $\widetilde{\Sigma}_{t+r}$ as reference information and evaluates the refined surface at arbitrary target coordinates $(m_i,\tau_i)$. This coordinate based construction allows data fitting terms to be evaluated at observed market locations and geometric residuals to be evaluated on synthetic collocation grids.

For a target coordinate $(m_i,\tau_i)$, SAAM applies learnable encoders $f_Q$, $f_K$, and $f_V$ to construct the query, key, and value embeddings:
\begin{equation}
\mathbf{Q}_i=f_Q(\mathbf{x}_i^Q), \qquad \mathbf{K}_j=f_K(\mathbf{x}_j^K), \qquad \mathbf{V}_j=f_V(\mathbf{x}_j^V).
\end{equation}
The cross sectional attention weights and the aggregated representation are computed by scaled dot product attention:
\begin{align}
\alpha_{ij} &= \frac{ \exp\left(\mathbf{Q}_i\mathbf{K}_j^{\top}/\sqrt{d}\right) }{ \sum_{n=1}^{N} \exp\left(\mathbf{Q}_i\mathbf{K}_n^{\top}/\sqrt{d}\right) }, \label{eq:attention_weights} \\
\mathbf{z}_i &= \sum_{j=1}^{N} \alpha_{ij}\mathbf{V}_j . \label{eq:attention_aggregation}
\end{align}
The refined implied volatility is then obtained through an output network and then a Softplus activation:
\begin{equation}
\widehat{\Sigma}_{\mathrm{ref},t+r}(m_i,\tau_i;\theta_{\mathrm{att}}) = \operatorname{Softplus} \left( f_{\mathrm{out}}(\mathbf{z}_i) \right).
\label{eq:saam_output}
\end{equation}
Thus, the refined value at each coordinate depends on the whole representative forecast surface rather than only on a local neighborhood.

\subsubsection{SAAM Variants}
We have two SAAM variants differ in how the query, key, and value inputs are constructed for cross sectional attention.
\begin{itemize}
\item \textbf{Spatial Separation SAAM.}
Attention routing is based on spatial coordinates, while volatility information enters through the value embedding:
\begin{equation}
\mathbf{x}_i^Q=(m_i,\tau_i), \qquad \mathbf{x}_j^K=(m_j,\tau_j), \qquad \mathbf{x}_j^V=\widetilde{\Sigma}_{t+r}(m_j,\tau_j).
\end{equation}
This design separates spatial routing from the predicted volatility magnitude and emphasizes the geometric organization of the moneyness--maturity domain.

\item \textbf{Volatility Fusion SAAM.}
Spatial inputs are augmented with volatility level information. The target query incorporates a global volatility level proxy, while the reference keys and values incorporate pointwise predicted volatility:
\begin{equation}
\mathbf{x}_i^Q=(m_i,\tau_i,\bar{v}_{t+r}), \qquad \mathbf{x}_j^K=\mathbf{x}_j^V = (m_j,\tau_j,\widetilde{\Sigma}_{t+r}(m_j,\tau_j)),
\end{equation}
where
\begin{equation}
\bar{v}_{t+r} = \frac{1}{N} \sum_{j=1}^{N} \widetilde{\Sigma}_{t+r}(m_j,\tau_j)
\end{equation}
is the average predicted volatility level of the representative forecast surface at horizon $r$. This design allows the attention weights to depend on both spatial geometry and the prevailing volatility level regime.
\end{itemize}

\begin{remark}
The global volatility level proxy is motivated by empirical evidence that IVS dynamics are often driven by low dimensional common factors, especially a dominant level factor \citep{Cont2002a}. In this framework, the proxy is not an additional no-arbitrage constraint; it serves as a regime level conditioning signal for cross sectional information routing.
\end{remark}

\subsubsection{Training Objective and Synthetic Collocation Grids}
\label{sec:reconstruction_and_arbitrage_loss}

SAAM is trained by combining data fitting on scattered market observations with penalties for the static no-arbitrage conditions in 
Proposition~\ref{prop:geometric_constraints}. Let $\mathcal{T}_{\mathrm{train}}$ denote the set of training forecast origins. 
For each $t\in\mathcal{T}_{\mathrm{train}}$ and $r\in\mathcal{R}_h$, define the total implied variance associated with the refined representative surface as
\begin{equation}
\phi_{\theta_{\mathrm{att}},t+r}(m,\tau) = \widehat{\Sigma}_{\mathrm{ref},t+r}(m,\tau;\theta_{\mathrm{att}})^2\tau .
\label{eq:refined_total_variance}
\end{equation}

The data fitting loss is evaluated on the scattered observation set $\Gamma_{t+r}$:
\begin{equation}
\mathcal{L}_{\mathrm{rec}} = \frac{1}{|\mathcal{T}_{\mathrm{train}}||\mathcal{R}_h|} \sum_{t\in\mathcal{T}_{\mathrm{train}}} \sum_{r\in\mathcal{R}_h} \frac{1}{n_{t+r}} \sum_{i=1}^{n_{t+r}} \left( \widehat{\Sigma}_{\mathrm{ref},t+r}(m_i,\tau_i;\theta_{\mathrm{att}}) - \Sigma_{t+r}(m_i,\tau_i) \right)^2,
\label{eq:rec_loss_quote}
\end{equation}
where $(m_i,\tau_i,\Sigma_{t+r}(m_i,\tau_i))\in\Gamma_{t+r}$. This term measures the discrepancy between the refined surface and the observed implied volatilities at available market locations.

The first three conditions are enforced by the architecture and therefore require no additional penalty terms. Conditions 1 and 2 are satisfied by the output parametrization: the output network $f_{\mathrm{out}}$, followed by the final Softplus layer, yields positive implied volatility values, and the definition $\phi_{\theta_{\mathrm{att}},t+r}(m,\tau) = \widehat{\Sigma}_{\mathrm{ref},t+r}(m,\tau;\theta_{\mathrm{att}})^2 \tau$ satisfies the zero-maturity boundary through the multiplicative factor $\tau$. Condition 3 is satisfied by construction, since the encoders, $f_{\mathrm{out}}$, and the final Softplus layer use twice-differentiable operations. Hence, the penalty terms associated with Conditions 1 to 3 are set to zero.
The calendar and butterfly penalties are evaluated by substituting $\phi=\phi_{\theta_{\mathrm{att}},t+r}$ into the residual operators $\ell_{\mathrm{cal}}$ and $\ell_{\mathrm{but}}$. With $[x]_{+}=\max(x,0)$, the corresponding penalty terms are
\begin{align}
\mathcal{L}_{\mathrm{cal}} &= \frac{1}{|\mathcal{T}_{\mathrm{train}}||\mathcal{R}_h|} \sum_{t\in\mathcal{T}_{\mathrm{train}}} \sum_{r\in\mathcal{R}_h} \frac{1}{|\mathcal{I}_{\mathrm{syn}}|} \sum_{(m_i,\tau_i)\in\mathcal{I}_{\mathrm{syn}}} \left[ -\ell_{\mathrm{cal}}\left(\phi_{\theta_{\mathrm{att}},t+r}\right)(m_i,\tau_i) \right]_{+}, \label{eq:loss_cal} \\
\mathcal{L}_{\mathrm{but}} &= \frac{1}{|\mathcal{T}_{\mathrm{train}}||\mathcal{R}_h|} \sum_{t\in\mathcal{T}_{\mathrm{train}}} \sum_{r\in\mathcal{R}_h} \frac{1}{|\mathcal{I}_{\mathrm{syn}}|} \sum_{(m_i,\tau_i)\in\mathcal{I}_{\mathrm{syn}}} \left[ -\ell_{\mathrm{but}}\left(\phi_{\theta_{\mathrm{att}},t+r}\right)(m_i,\tau_i) \right]_{+}.
\label{eq:loss_but}
\end{align}

For the tail condition, a global check as $|m|\to\infty$ is not possible on a finite empirical domain. We therefore use a boundary curvature penalty as a finite grid proxy:
\begin{equation}
\mathcal{L}_{\mathrm{wing}} = \frac{1}{|\mathcal{T}_{\mathrm{train}}||\mathcal{R}_h|} \sum_{t\in\mathcal{T}_{\mathrm{train}}} \sum_{r\in\mathcal{R}_h} \frac{1}{|\mathcal{I}_{\mathrm{wing}}|} \sum_{(m_i,\tau_i)\in\mathcal{I}_{\mathrm{wing}}} \left| \frac{\partial^2 \phi_{\theta_{\mathrm{att}},t+r}(m_i,\tau_i)} {\partial m^2} \right|.
\label{eq:loss_wing}
\end{equation}

The residuals are evaluated on dense synthetic grids over moneyness and maturity. Set $m_{\min}=\log(0.6)$, $m_{\max}=\log(1.5)$, and $\tau_{\max}=1$. The synthetic grids are
\begin{align}
\mathcal{M}_{\mathrm{syn}} &= \left\{ x^3 \mid x\in \operatorname{linspace} \left( -(-2m_{\min})^{1/3}, (2m_{\max})^{1/3}, 60 \right) \right\}, \\
\mathcal{T}_{\mathrm{syn}} &= \left\{ \exp(x) \mid x\in \operatorname{linspace} \left( \log(1/365), \log(\tau_{\max}), 60 \right) \right\}.
\end{align}

The calendar and butterfly penalties are evaluated on
\begin{equation}
\mathcal{I}_{\mathrm{syn}} = \mathcal{M}_{\mathrm{syn}}\times\mathcal{T}_{\mathrm{syn}},
\end{equation}
whereas $\mathcal{I}_{\mathrm{wing}}$ consists of boundary moneyness points selected from $\mathcal{M}_{\mathrm{syn}}$ across the maturity grid. The same synthetic grid is used for out-of-sample finite-grid static no-arbitrage residual diagnostics, so that $\mathcal{I}_{\mathrm{eval}}=\mathcal{I}_{\mathrm{syn}}$.
The total SAAM objective is
\begin{equation}
\mathcal{L}_{\mathrm{SAAM}} = \mathcal{L}_{\mathrm{rec}} + \lambda \left( \mathcal{L}_{\mathrm{cal}} + \mathcal{L}_{\mathrm{but}} + \mathcal{L}_{\mathrm{wing}} \right).
\label{eq:saam_total_loss}
\end{equation}
The resulting objective combines empirical data fitting with residual regularization for the main static no-arbitrage diagnostics. It improves fit to observed implied 
volatilities while reducing finite grid residual violations on the inspected diagnostic grid, but it is not an exact projection onto the continuous arbitrage free set.
\begin{remark}
The static no-arbitrage residuals are defined on a continuous moneyness and maturity domain, but they cannot be evaluated at every point in implementation. We therefore use $\mathcal{I}_{\mathrm{syn}}$ as a dense diagnostic grid for the calendar and butterfly residuals. The tail condition is asymptotic and cannot be verified on a finite empirical domain, so it is examined through the boundary set $\mathcal{I}_{\mathrm{wing}}$. Consequently, small residual values should be interpreted as numerical evidence that the refined surface satisfies the inspected finite domain diagnostics, rather than as a formal proof of global static arbitrage freedom.
\end{remark}

\section{Empirical Analysis and Results}
\label{sec:experiments}

This section evaluates the proposed decoupled framework using CSI 300 index options. Consistent with the two stage methodology, we first examine the fixed grid 
predictive accuracy and distributional forecast quality of the diffusion stage. We then assess whether SAAM improves fit to market observations and reduces 
static no-arbitrage residual violations on the diagnostic grid. The section concludes with attention diagnostics that describe how cross sectional information 
is allocated across the surface.

\subsection{Experimental Setup}
\label{sec:experimental_setup}

\subsubsection{Data Characteristics and Empirical Irregularities}
\label{sec:exp_dataset}

We use CSI 300 index option data from RiceQuant, covering the period from June 10, 2020, to September 30, 2024. Implied volatilities are computed from end of interval option prices: daily observations use end of day prices, while minute level observations use the last available price within each one minute interval. The data are processed following the construction procedure in Section~\ref{sec:data_grid}. Specifically, the original scattered market observations are first mapped onto fixed grid IVS states for Stage I forecasting, and are then used directly as observation targets in Stage II refinement. The main descriptive evidence in this subsection is reported at the daily frequency, while the corresponding minute level diagnostics are provided in Appendix~\ref{app:minute level_violations}.

\begin{table*}[htbp]
    \centering
    \caption{Descriptive statistics of implied volatility for CSI 300 index options, June 2020--September 2024.}
    \label{tab:iv_descriptive_stats}
    \small
    \setlength{\tabcolsep}{5pt}
    \renewcommand{\arraystretch}{1.12}
    \begin{tabular}{lrrrrrrrr}
    \hline
    Category & N & Mean & Std & Skewness & Kurtosis & P5 & Median & P95 \\
    \hline
    \multicolumn{9}{l}{Panel A: By moneyness} \\
    \hline
    Deep OTM Put  & 18,139 & 0.3299 & 0.1806 & 3.2734 & 15.4377 & 0.1814 & 0.2792 & 0.6571 \\
    OTM Put       & 63,925 & 0.2071 & 0.0678 & 2.2701 & 15.0951 & 0.1176 & 0.1989 & 0.3082 \\
    ATM           & 40,479 & 0.1899 & 0.0492 & 1.1812 & 4.0362  & 0.1278 & 0.1794 & 0.2846 \\
    OTM Call      & 64,588 & 0.2236 & 0.0772 & 2.1906 & 9.7272  & 0.1417 & 0.2030 & 0.3619 \\
    Deep OTM Call & 19,430 & 0.3209 & 0.1572 & 2.6977 & 12.3670 & 0.1782 & 0.2730 & 0.6096 \\
    \hline
    \multicolumn{9}{l}{Panel B: By time to maturity} \\
    \hline
    Short, $\tau\leq 60$ days                & 85,461 & 0.2606 & 0.1401 & 3.4229 & 19.6807 & 0.1410 & 0.2231 & 0.5112 \\
    Medium, $61\leq \tau\leq 180$ days       & 71,701 & 0.2134 & 0.0672 & 1.4925 & 4.9278  & 0.1302 & 0.2003 & 0.3356 \\
    Long, $\tau>180$ days                    & 49,399 & 0.2027 & 0.0561 & 0.6609 & 0.6770  & 0.1258 & 0.1941 & 0.3019 \\
    \hline
    All & 206,561 & 0.2304 & 0.1053 & 3.9990 & 30.9768 & 0.1331 & 0.2067 & 0.3992 \\
    \hline
    \multicolumn{9}{p{14.2cm}}{\footnotesize \textit{Note:} P5 and P95 denote the 5th and 95th percentiles. Skewness and kurtosis are computed using pooled observations within each moneyness or maturity regime.}
    \end{tabular}
\end{table*}

Table~\ref{tab:iv_descriptive_stats} reports summary statistics of daily implied volatility across moneyness and maturity regimes. The sample displays strong positive 
skewness and high kurtosis, especially in short maturity and deep out-of-the-money regions. These patterns indicate that the conditional distribution of future IVS states 
is far from Gaussian and may exhibit pronounced tail behavior. This motivates the use of a probabilistic scenario generation model rather than a purely point predictor.

\begin{figure*}[htbp!]
\centering
\includegraphics[width=0.85\textwidth]{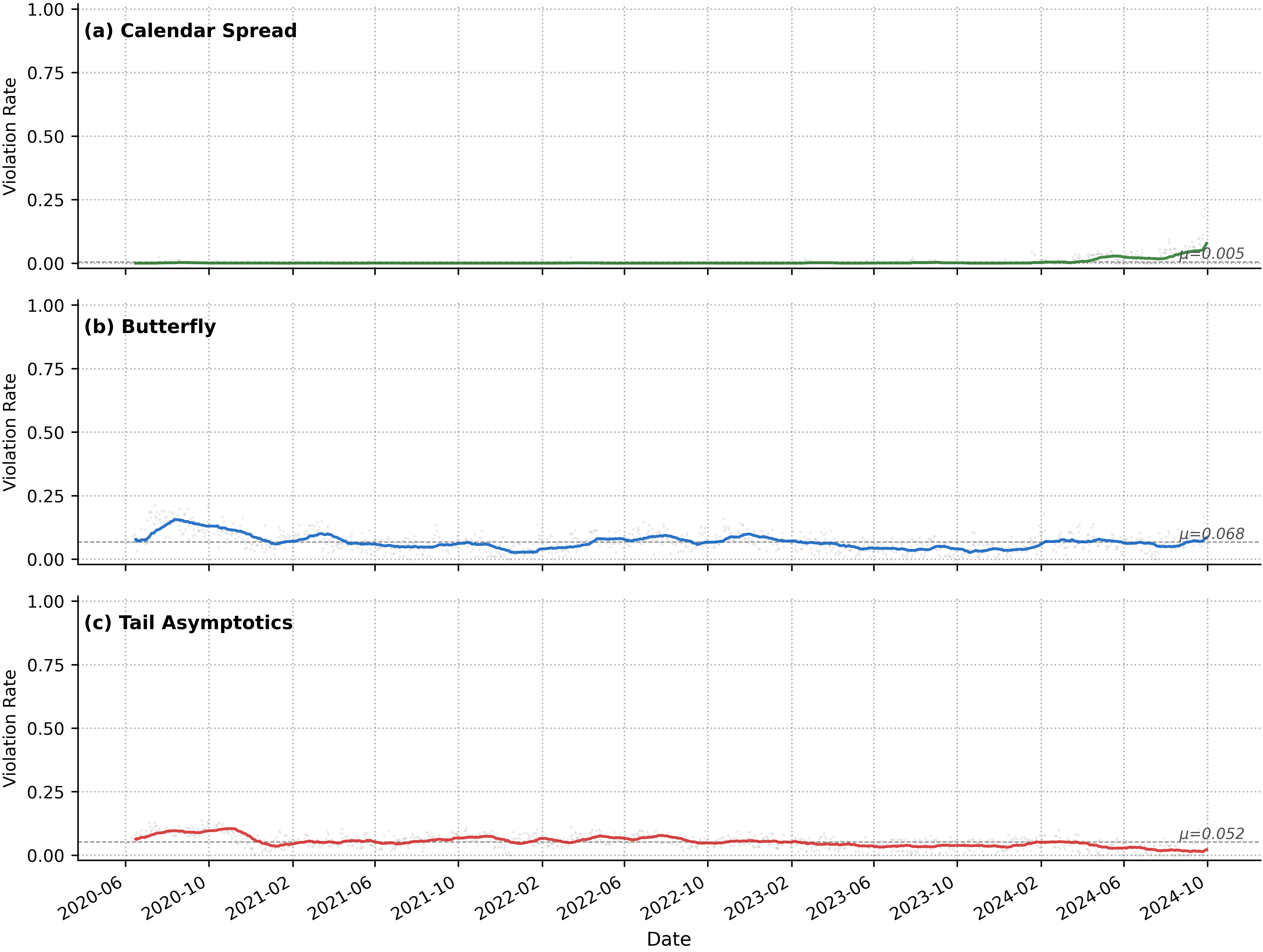}
\caption{Temporal persistence of static no-arbitrage residual violations in CSI 300 index options based on raw daily observations.}
\label{fig:arbitrage_analysis}
\end{figure*}

Figure~\ref{fig:arbitrage_analysis} shows that static no-arbitrage residual violations are not merely isolated outliers. Calendar spread deviations remain relatively limited, 
while butterfly spread deviations are more pronounced and persistent over time. The butterfly spread residual is particularly important because it reflects strike wise 
convexity and is linked to the admissibility of the implied risk neutral density. Together with the distributional evidence in Table~\ref{tab:iv_descriptive_stats}, these results show that 
the data contain both stochastic complexity and static no-arbitrage violations. This motivates the decoupled empirical design: Stage I models conditional IVS scenarios 
on the fixed grid, while Stage II refines a representative forecast surface to improve fit to market observations and reduce static no-arbitrage violations.

\subsubsection{Forecasting Protocols, Baselines, and Implementation}
\label{sec:baselines}
We evaluate the proposed framework under two forecasting protocols:
\begin{itemize}
\item \textbf{Daily forecasting.} A rolling window of $H=22$ trading days is used to predict the next-day fixed-grid IVS state, with $\mathcal{H}_{\mathrm{eval}}^{D}=\{1\}$. The daily sample is split chronologically into 80\% training and 20\% testing observations.
\item \textbf{Minute Level forecasting.} The previous 60 minutes of IVS states are used to forecast the next five minute-level horizons, with $\mathcal{H}_{\mathrm{eval}}^{I}=\{1,\ldots,5\}$. The minute level sample is divided into 70\% training, 10\% validation, and 20\% testing observations.
\end{itemize}
In what follows, $\mathcal{H}_{\mathrm{eval}}$ denotes the relevant evaluation horizon set under the daily or minute level protocol.

For Stage I, the conditional diffusion model is compared numerically with LSTM-Direct on the fixed grid $\mathcal{D}_0$, where LSTM-Direct serves as the main  
recurrent benchmark. LSTM-VAE \citep{Zhang2022} is included as a latent variable benchmark in the graphical analysis, since its sampled output is not based on 
the same  evaluation protocol. For each test origin, the diffusion model generates $B=100$ scenarios. The full ensemble is used for probabilistic 
evaluation, while the pointwise median $\widehat{\Sigma}_{t+r}^{\mathrm{med}}$ serves as a representative surface for  comparison and Stage-II refinement.

For Stage II, we compare Spatial Separation SAAM and Volatility Fusion SAAM with MLP based refinement pipelines \citep{Ackerer2020,Zheng2021,Zhang2022}. The MLP 
baselines follow recurrent forecasting pipelines and use pointwise refinement mappings, whereas the two SAAM variants refine the diffusion median within the 
proposed framework. All final refined surfaces are evaluated at the same market observation coordinates and on the same static no-arbitrage diagnostic grid. 
Therefore, the comparison is conducted at the pipeline level under a common evaluation protocol, rather than as an input controlled ablation of refinement operators.

The main implementation settings are as follows. The diffusion model uses a Transformer based score network with 64 channels, 8 attention heads, and 
4 residual blocks. SAAM uses latent dimension $d=64$, with 4 cross attention heads for Volatility Fusion SAAM and 3 for Spatial Separation SAAM. All 
models are implemented in PyTorch and trained with the Adam optimizer on a single NVIDIA RTX 4090 GPU. Penalty weights are selected by grid search, 
with sensitivity results reported in Appendix~\ref{app:penalty_weight_selection}.

A direct numerical comparison with VolGAN \citep{Vuletić2025} is not included, because VolGAN is designed for joint scenario simulation of the underlying 
asset and the implied volatility surface and handles static no-arbitrage considerations through ex-post scenario reweighting. This reweighting changes the 
probability weights over generated scenarios rather than directly modifying a given surface. Our evaluation focuses instead on conditional IVS forecasting 
and surface level refinement with static no arbitrage residual control.

\subsubsection{Evaluation Metrics}
\label{sec:metrics}
We evaluate the framework from three perspectives: forecasting accuracy, fit to market observations, and finite grid static no-arbitrage residual 
diagnostics. Let $\mathcal{T}_{\mathrm{test}}$ denote the set of out-of-sample forecast origins and $\mathcal{H}_{\mathrm{eval}}$ the evaluation horizon set 
under the daily or minute-level protocol. For any scalar $x$, write $[x]_{+}=\max(x,0)$.

For Stage I, accuracy is evaluated on the fixed grid $\mathcal{D}_0$. For a forecasting method $M$, let $\widehat{\Sigma}_{t+r}^{M}$ denote its unrefined 
representative forecast surface at horizon $r$. For the diffusion model, $\widehat{\Sigma}_{t+r}^{M}$ is the pointwise ensemble median in Eq.~\eqref{eq:ensemble_median}. We report
\begin{align}
\mathrm{RMSE}_{\mathrm{grid}}(M)
&=
\left[
\frac{1}{|\mathcal{T}_{\mathrm{test}}||\mathcal{H}_{\mathrm{eval}}|}
\sum_{t\in\mathcal{T}_{\mathrm{test}}}
\sum_{r\in\mathcal{H}_{\mathrm{eval}}}
\frac{1}{|\mathcal{D}_0|}
\sum_{(m,\tau)\in\mathcal{D}_0}
\left(
\widehat{\Sigma}_{t+r}^{M}(m,\tau) - \bar{\Sigma}_{t+r}(m,\tau)
\right)^2
\right]^{1/2}, \\
\mathrm{MAE}_{\mathrm{grid}}(M)
&=
\frac{1}{|\mathcal{T}_{\mathrm{test}}||\mathcal{H}_{\mathrm{eval}}|}
\sum_{t\in\mathcal{T}_{\mathrm{test}}}
\sum_{r\in\mathcal{H}_{\mathrm{eval}}}
\frac{1}{|\mathcal{D}_0|}
\sum_{(m,\tau)\in\mathcal{D}_0}
\left|
\widehat{\Sigma}_{t+r}^{M}(m,\tau) - \bar{\Sigma}_{t+r}(m,\tau)
\right|.
\end{align}
For the diffusion model, distributional forecast quality is additionally measured by the Continuous Ranked Probability Score (CRPS), averaged over forecast origins, horizons, and grid points.

For Stage II, fitting accuracy is evaluated at the observed market locations in $\Gamma_{t+r}$. For a refinement method $M$, let $\widehat{\Sigma}_{\mathrm{ref},t+r}^{M}$ denote the refined surface at horizon $r$. We report
\begin{align}
\mathrm{RMSE}_{\mathrm{obs}}(M)
&=
\left[
\frac{1}{|\mathcal{T}_{\mathrm{test}}||\mathcal{H}_{\mathrm{eval}}|}
\sum_{t\in\mathcal{T}_{\mathrm{test}}}
\sum_{r\in\mathcal{H}_{\mathrm{eval}}}
\frac{1}{n_{t+r}}
\sum_{i=1}^{n_{t+r}}
\left(
\widehat{\Sigma}_{\mathrm{ref},t+r}^{M}(m_i,\tau_i) - \Sigma_{t+r}(m_i,\tau_i)
\right)^2
\right]^{1/2}, \\
\mathrm{MAPE}_{\mathrm{obs}}(M)
&=
\frac{1}{|\mathcal{T}_{\mathrm{test}}||\mathcal{H}_{\mathrm{eval}}|}
\sum_{t\in\mathcal{T}_{\mathrm{test}}}
\sum_{r\in\mathcal{H}_{\mathrm{eval}}}
\frac{1}{n_{t+r}}
\sum_{i=1}^{n_{t+r}}
\left|
\frac{
\widehat{\Sigma}_{\mathrm{ref},t+r}^{M}(m_i,\tau_i) - \Sigma_{t+r}(m_i,\tau_i)
}{
\Sigma_{t+r}(m_i,\tau_i)
}
\right|,
\end{align}
where $(m_i,\tau_i,\Sigma_{t+r}(m_i,\tau_i))\in\Gamma_{t+r}$.

Static no arbitrage residual violations are assessed using the average negative parts of the calendar spread and butterfly spread residuals on the dense evaluation grid. For a refined surface produced by method $M$, define
\[
\phi_{\mathrm{ref},t+r}^{M}(m,\tau) = \left( \widehat{\Sigma}_{\mathrm{ref},t+r}^{M}(m,\tau) \right)^2\tau .
\]
Substituting $\phi=\phi_{\mathrm{ref},t+r}^{M}$ into the calendar and butterfly residual operators gives $\ell_{\mathrm{cal},t+r}^{M}$ and $\ell_{\mathrm{but},t+r}^{M}$. We compute
\begin{align}
L_{\mathrm{cal}}^{-}(M)
&=
\frac{1}{|\mathcal{T}_{\mathrm{test}}||\mathcal{H}_{\mathrm{eval}}|}
\sum_{t\in\mathcal{T}_{\mathrm{test}}}
\sum_{r\in\mathcal{H}_{\mathrm{eval}}}
\frac{1}{|\mathcal{I}_{\mathrm{eval}}|}
\sum_{(m,\tau)\in\mathcal{I}_{\mathrm{eval}}}
\left[ -\ell_{\mathrm{cal},t+r}^{M}(m,\tau) \right]_{+},
\label{eq:violation_calendar} \\
L_{\mathrm{but}}^{-}(M)
&=
\frac{1}{|\mathcal{T}_{\mathrm{test}}||\mathcal{H}_{\mathrm{eval}}|}
\sum_{t\in\mathcal{T}_{\mathrm{test}}}
\sum_{r\in\mathcal{H}_{\mathrm{eval}}}
\frac{1}{|\mathcal{I}_{\mathrm{eval}}|}
\sum_{(m,\tau)\in\mathcal{I}_{\mathrm{eval}}}
\left[ -\ell_{\mathrm{but},t+r}^{M}(m,\tau) \right]_{+}.
\label{eq:violation_butterfly}
\end{align}
The grid $\mathcal{I}_{\mathrm{eval}}$ is identical to the synthetic collocation grid $\mathcal{I}_{\mathrm{syn}}$ used in the refinement loss. A zero value of $L_{\mathrm{cal}}^{-}(M)$ or $L_{\mathrm{but}}^{-}(M)$ indicates that no corresponding violation is detected on this finite evaluation grid.

\subsection{Stage-I Forecasting Performance}
\label{sec:prob_forecasting}

We first evaluate the conditional diffusion stage. The evaluation is conducted on the  $\mathcal{D}_0$ and considers both the representative point forecast, 
given by the pointwise median of the predictive ensemble, and the distributional information contained in the generated scenarios.

\begin{table}[htbp]
\centering
\setlength{\tabcolsep}{6pt}
\renewcommand{\arraystretch}{1.15}
\caption{Out-of-sample Stage-I forecasting performance. RMSE and MAE are computed from representative point forecasts; CRPS is reported only for the diffusion ensemble. Results are mean $\pm$ standard deviation over five independent runs.}
\label{tab:performance_comparison}
\begin{tabular}{lcccc}
\hline
Metric & \multicolumn{2}{c}{Daily Forecasting} & \multicolumn{2}{c}{Minute Level Forecasting} \\
\hline
& Conditional Diffusion & LSTM-Direct & Conditional Diffusion & LSTM-Direct \\
\hline
RMSE & $\mathbf{0.0263} \pm \mathbf{0.0006}$ & $0.0288 \pm 0.0007$ & $\mathbf{0.0114} \pm \mathbf{0.0004}$ & $0.0178 \pm 0.0001$ \\
MAE & $\mathbf{0.0129} \pm \mathbf{0.0002}$ & $0.0176 \pm 0.0008$ & $\mathbf{0.0017} \pm \mathbf{0.0001}$ & $0.0110 \pm 0.0001$ \\
CRPS & $\mathbf{0.0350} \pm \mathbf{0.0018}$ & N/A & $\mathbf{0.0078} \pm \mathbf{0.0005}$ & N/A \\
\hline
\end{tabular}
\end{table}

Table~\ref{tab:performance_comparison} shows that the diffusion median improves  point-forecast accuracy relative to LSTM-Direct. 
The RMSE reduction is approximately 8.7\% in the daily experiment and 36.0\% in the minute level experiment. The larger gain at the minute level 
frequency suggests that the diffusion model is particularly effective when the IVS evolves with high-frequency noise and short-term local fluctuations. 
The CRPS values further indicate that the generated ensemble provides useful distributional information beyond its median forecast.

\begin{figure*}[p]
\centering
\begin{minipage}{0.82\textwidth}
\centering
\includegraphics[width=\linewidth]{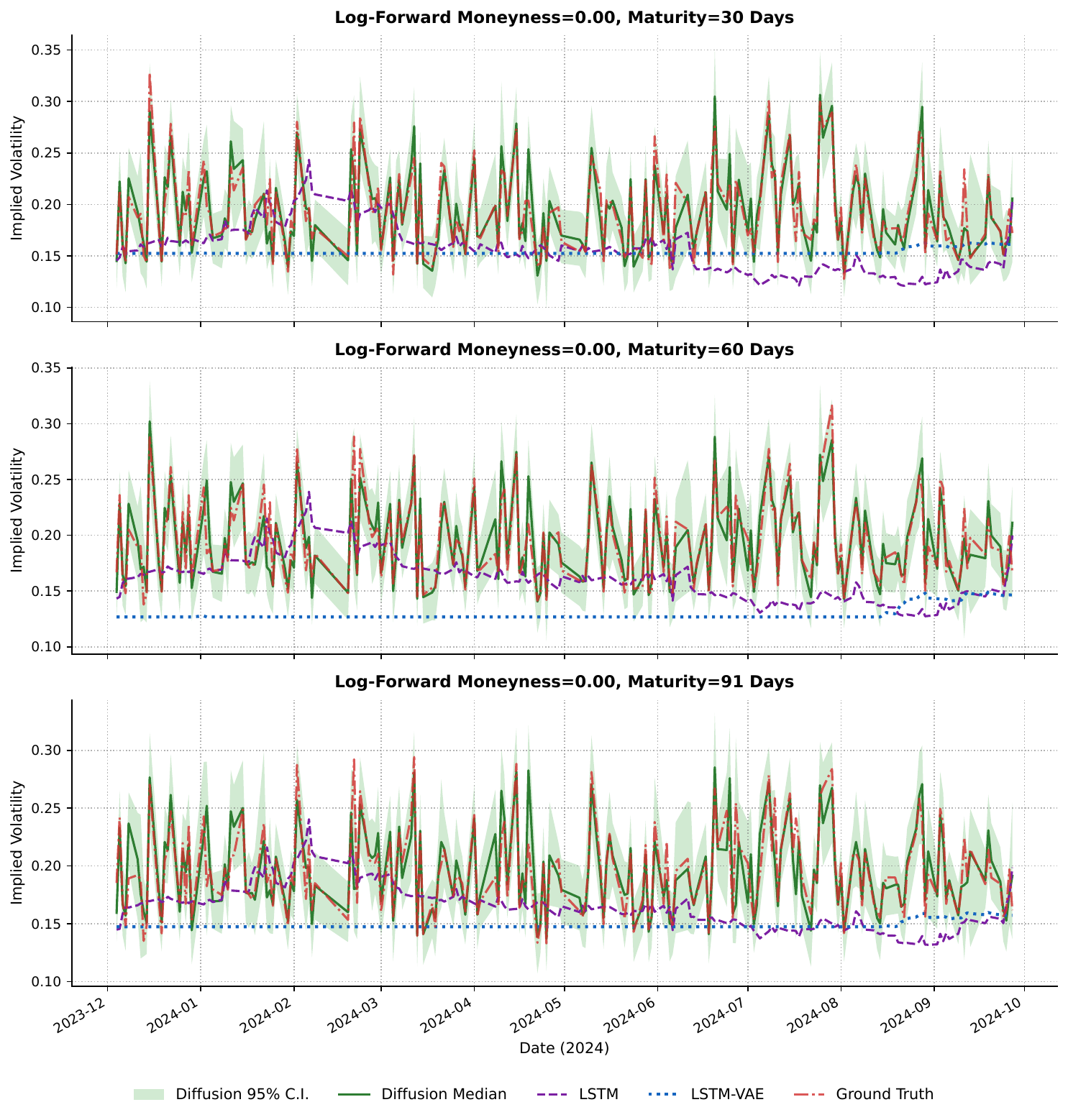}
\centerline{\small (a) At-the-money coordinates across maturities.}
\end{minipage}
\vspace{0.35cm}
\begin{minipage}{0.82\textwidth}
\centering
\includegraphics[width=\linewidth]{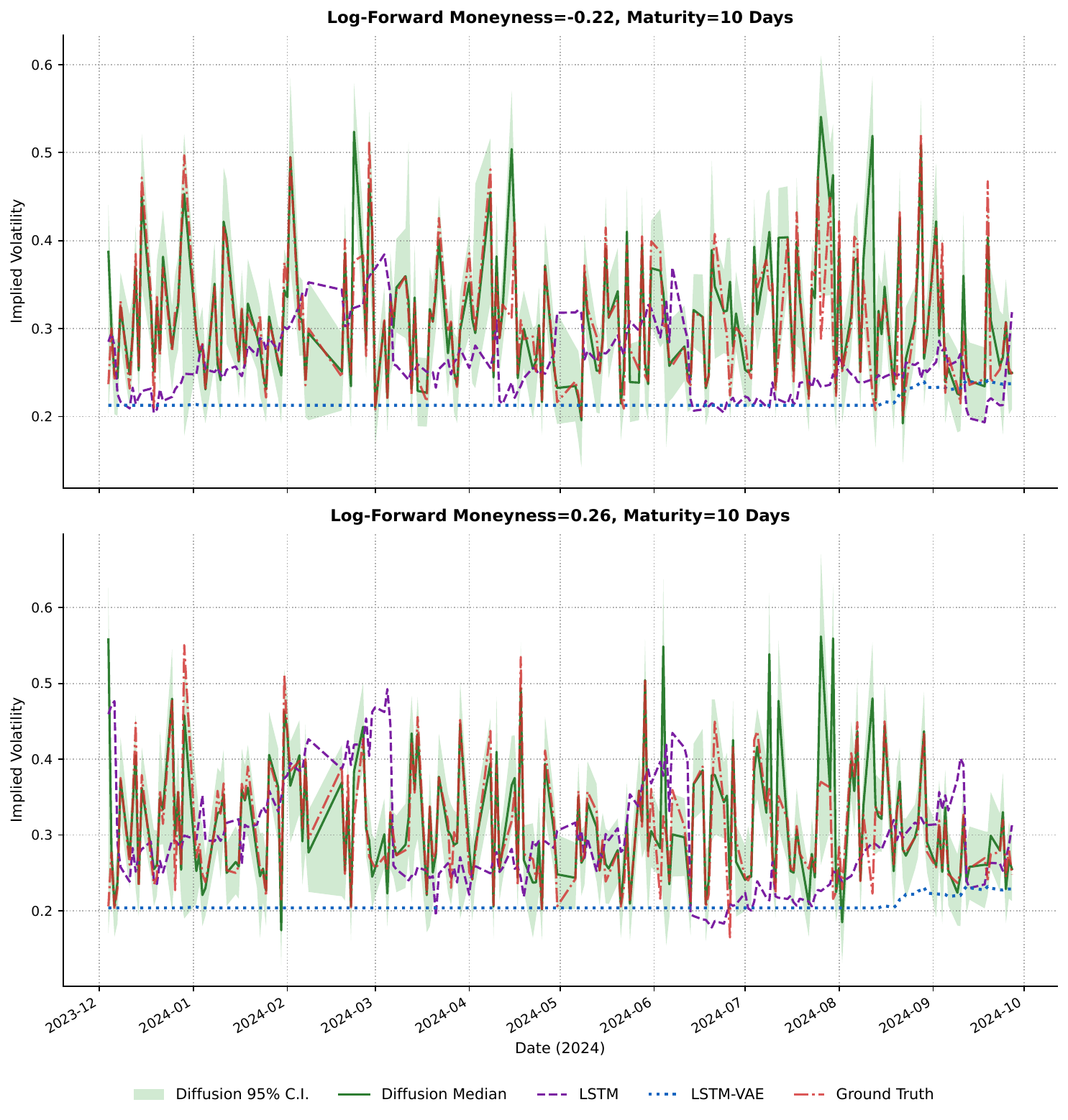}
\centerline{\small (b) Short-maturity deep out-of-the-money coordinates.}
\end{minipage}
\vspace{0.15cm}
\caption{Out-of-sample daily forecasting trajectories at representative IVS coordinates. Panel (a) reports at-the-money trajectories for maturities of 30, 60, and 91 days. Panel (b) reports short-maturity deep out-of-the-money trajectories for $m=-0.22$ and $m=0.26$, with $\tau=10$ days. The shaded region denotes the 95\% predictive interval of the conditional diffusion ensemble.}
\label{fig:daily_forecasting_combined}
\end{figure*}

Figure~\ref{fig:daily_forecasting_combined} provides qualitative evidence consistent with the numerical results. At the ATM coordinates (Panel a), the diffusion median follows the 
implied volatility trajectory across maturities, while the predictive interval expands during periods of larger implied volatility movements. In the deep OTM regions 
(Panel b), the trajectories are more irregular and the diffusion intervals are wider, reflecting greater uncertainty in sparse and tail-sensitive parts of the surface. 
By contrast, the LSTM-VAE trajectories are nearly time-invariant in both panels, indicating that the latent-variable reconstruction tends to smooth away short-term surface dynamics.

\begin{figure*}[htbp]
\centering
\begin{minipage}{0.9\textwidth}
\centering
\includegraphics[width=\linewidth]{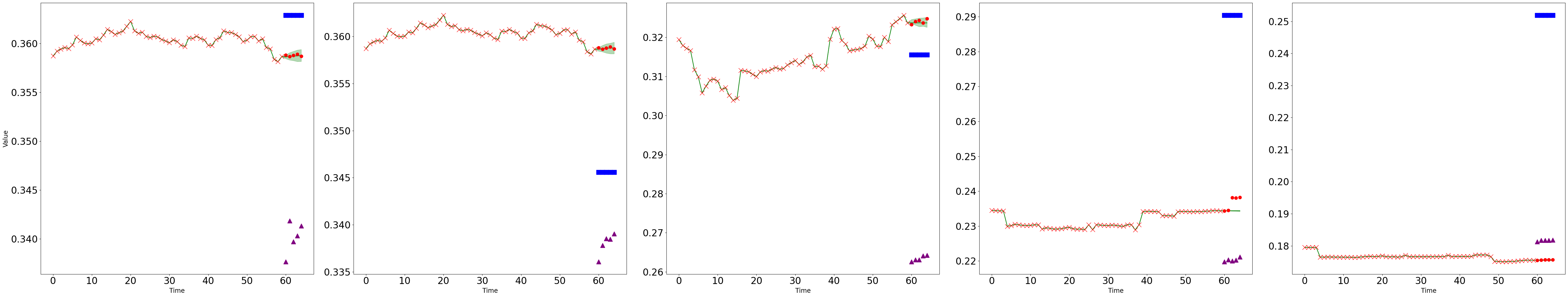}
\centerline{\small (a) Typical high-frequency fluctuations.}
\end{minipage}
\vspace{0.3cm}
\begin{minipage}{0.9\textwidth}
\centering
\includegraphics[width=\linewidth]{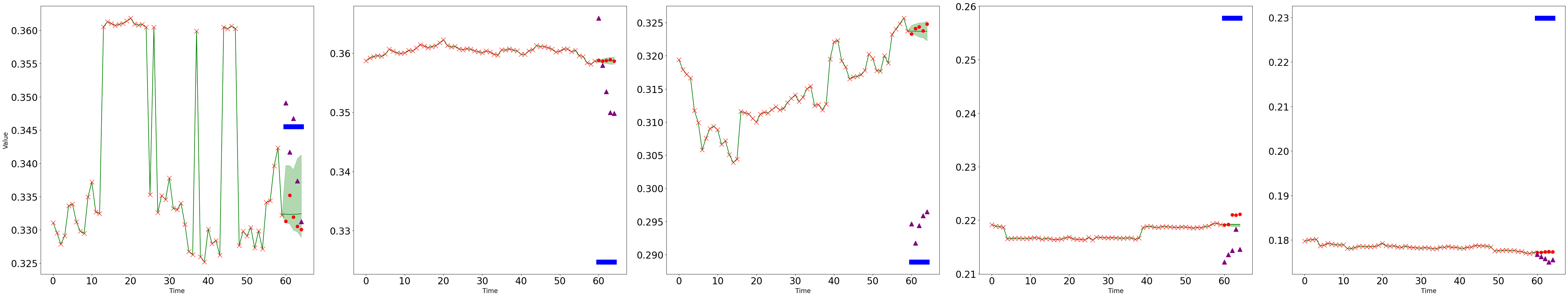}
\centerline{\small (b) Abrupt market regime shifts.}
\end{minipage}
\vspace{0.2cm}
\caption{Minute-level forecasting trajectories. Red markers denote observed IV values; the green line and shaded band denote the diffusion median and predictive interval; purple and blue markers denote LSTM-Direct and LSTM-VAE forecasts, respectively.}
\label{fig:minute level_combined}
\end{figure*}

Figure~\ref{fig:minute level_combined} shows the same pattern in the minute-level experiment. The diffusion forecasts remain responsive to short-term movements and provide uncertainty bands around the observed IV  trajectory. The LSTM-Direct forecasts exhibit larger displacement in several horizons, while the LSTM-VAE forecasts again remain close to a flat trajectory. This behavior is consistent with the latent bottleneck of VAE-type sequence models, which can suppress local variation when reconstructing high-frequency IVS dynamics.

Overall, the Stage I results show that conditional diffusion is effective for IVS scenario generation. It improves  forecast accuracy relative 
to  recurrent baselines and produces predictive intervals that reflect heterogeneous dispersion across moneyness, maturity, and sampling frequency. 
These results support the scenario generation stage of the proposed framework. However, predictive accuracy and distributional flexibility do not by themselves 
control static no-arbitrage residual violations, which motivates the Stage II surface refinement step.

\subsection{Stage-II Refinement and Static No-Arbitrage Diagnostics}
\label{sec:refinement_results}

We next evaluate the Stage II refinement module. It takes the robust representative 
diffusion median as input and refines it into a surface that better fits market observations while improving  static no-arbitrage residual control. 
We compare the two SAAM variants with MLP refinement baselines under both daily and minute level settings.

\subsubsection{Fitting Accuracy and Residual Violations}

Table~\ref{tab:SAAM_vs_DNN_Comprehensive} reports the numerical performance of the refinement stage. RMSE and MAPE evaluate fitting accuracy at observed market 
locations, while $L_{\mathrm{cal}}^{-}$ and $L_{\mathrm{but}}^{-}$ measure the average negative parts of the calendar-spread and butterfly-spread residuals on the 
synthetic evaluation grid. These two residual metrics diagnose violations of term-structure monotonicity and strike-convexity admissibility, respectively.

\begin{table*}[htbp]
\centering
\setlength{\tabcolsep}{7pt}
\renewcommand{\arraystretch}{1.15}
\caption{Numerical evaluation of the Stage-II refinement module. Values are reported as mean $\pm$ standard deviation over five independent runs. RMSE and MAPE evaluate fitting accuracy against market observations, while $L_{\mathrm{cal}}^{-}$ and $L_{\mathrm{but}}^{-}$ measure average residual violations on the synthetic evaluation grid.}
\label{tab:SAAM_vs_DNN_Comprehensive}
\begin{tabular}{lcccc}
\hline
Model & Fusion SAAM & Separation SAAM & MLP-Direct & MLP-VAE \\
\hline
\multicolumn{5}{l}{Panel A: Daily regime} \\
\hline
RMSE & $\mathbf{0.0190} \pm \mathbf{0.0009}$ & $0.0217 \pm 0.0014$ & $0.0290 \pm 0.0026$ & $0.0341 \pm 0.0012$ \\
MAPE & $8.47\% \pm 0.40\%$ & $\mathbf{8.43\%} \pm \mathbf{0.59\%}$ & $10.16\% \pm 1.02\%$ & $13.99\% \pm 0.96\%$ \\
$L_{\mathrm{cal}}^{-}$ & $0.000$ & $0.0000$ & $0.0000$ & $0.0000$ \\
$L_{\mathrm{but}}^{-}$ & $0.000$ & $0.0000$ & $0.0000$ & $0.0000$ \\
\hline
\multicolumn{5}{l}{Panel B: Minute Level regime} \\
\hline
RMSE & $\mathbf{0.0215} \pm \mathbf{0.0009}$ & $0.0329 \pm 0.0081$ & $0.0830 \pm 0.0008$ & $0.1015 \pm 0.0007$ \\
MAPE & $\mathbf{9.54\%} \pm \mathbf{0.95\%}$ & $12.47\% \pm 1.32\%$ & $17.87\% \pm 0.23\%$ & $19.37\% \pm 0.12\%$ \\
$L_{\mathrm{cal}}^{-}$ & $0.0000$ & $0.0000$ & $0.0000$ & $0.0000$ \\
$L_{\mathrm{but}}^{-}$ & $0.0002 \pm 0.0001$ & $0.0006 \pm 0.0002$ & $0.0000$ & $0.0000$ \\
\hline
\end{tabular}
\end{table*}

In the daily setting, all methods exhibit no detected calendar spread or butterfly spread violations on the evaluation grid. The main distinction lies in fit to 
market observations. Volatility Fusion SAAM achieves the lowest RMSE, while Spatial Separation SAAM achieves the lowest MAPE. Both SAAM variants outperform the 
MLP refiners, suggesting that cross-sectional attention improves fitting accuracy without weakening  static no-arbitrage diagnostics.

The minute level regime is more demanding. The MLP refiners report zero butterfly-spread residuals on the inspected grid, but this comes with substantially larger RMSE and MAPE. This indicates that eliminating residual violations on a finite grid is not sufficient if the resulting surface loses too much agreement with observed market locations. By contrast, Fusion SAAM achieves the best minute level RMSE and MAPE while keeping $L_{\mathrm{but}}^{-}$ close to zero. The penalty-weight sensitivity reported in Appendix~\ref{app:penalty_weight_selection} provides additional context: stronger geometric penalization can reduce residual violations but may also degrade fitting accuracy. The results in Table~\ref{tab:SAAM_vs_DNN_Comprehensive} show that SAAM attains a more favorable fidelity--regularity trade-off than the pointwise MLP refiners.

Overall, the quantitative evidence supports the decoupled design. Stage I captures stochastic fixed grid IVS dynamics, while Stage II refines the representative 
forecast surface to improve fit to market observations and strengthen static no-arbitrage residual control. The advantage of SAAM is particularly clear in the 
minute level setting, where local noise and sparse observations make pointwise correction less stable.

\subsubsection{Visualization of Refined Implied Volatility Surfaces}

\begin{figure*}[htbp]
\centering
\includegraphics[width=1.0\textwidth]{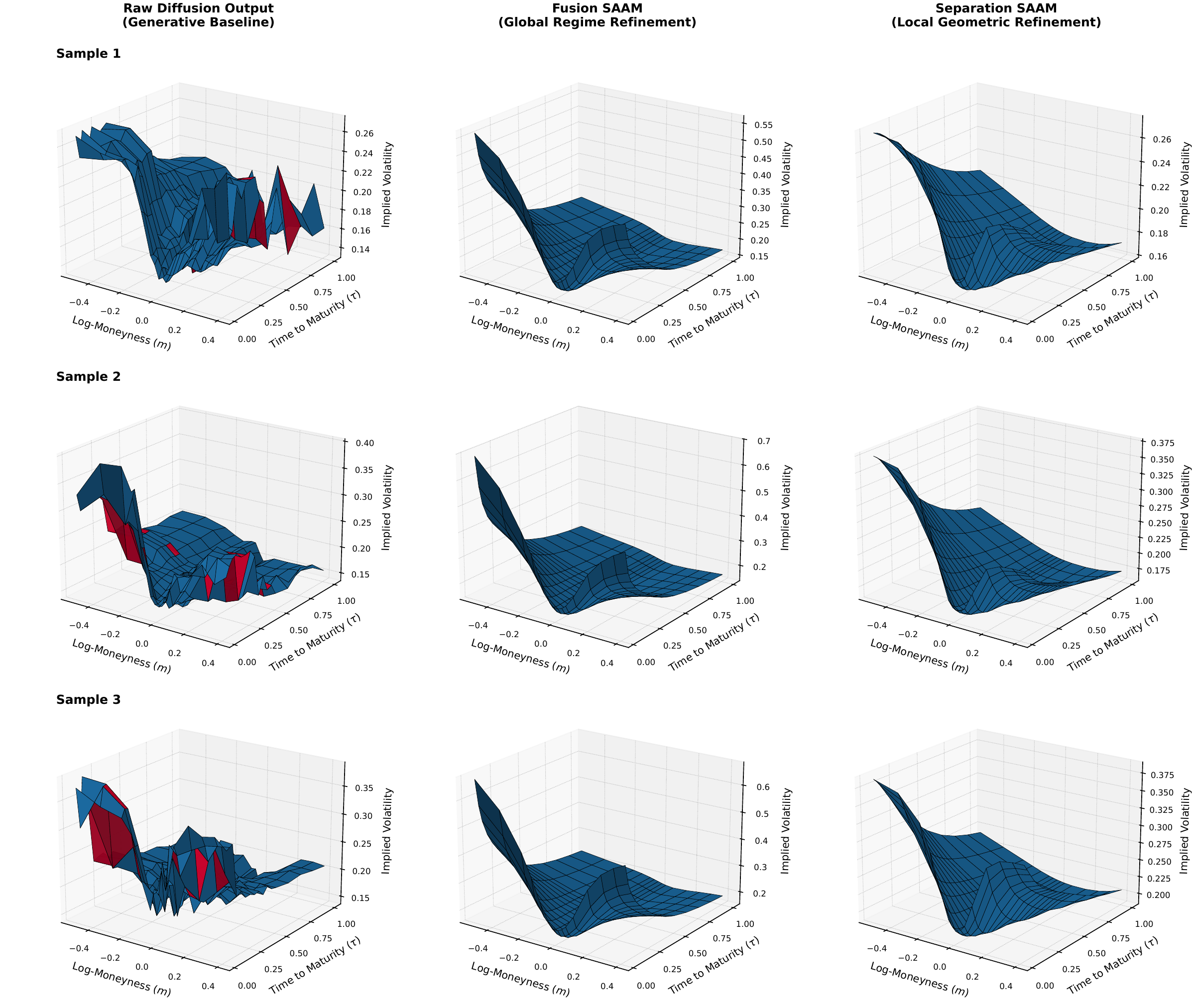}
\caption{Visual comparison of Stage II surface refinement on representative daily samples. The first column shows unrefined diffusion outputs, where red regions indicate localized static no arbitrage residual violations on the diagnostic grid. The second column reports Volatility Fusion SAAM, and the third column reports Spatial Separation SAAM.}
\label{fig:ablation_3x3}
\end{figure*}

Figure~\ref{fig:ablation_3x3} illustrates how the refinement module changes the representative diffusion surface. The unrefined diffusion outputs capture 
the broad cross-sectional shape of the IVS, but they may contain localized violation regions on the diagnostic grid, especially in less regular parts of 
the moneyness--maturity domain. This confirms that accurate fixed grid forecasting alone does not necessarily produce a surface satisfying the no-arbitrage diagnostics.

The two SAAM variants reduce these localized violations through different refinement patterns. Volatility Fusion SAAM produces smoother and more globally 
coordinated adjustments, consistent with its use of a volatility level conditioning signal. Spatial Separation SAAM makes more localized geometric corrections 
and tends to preserve more of the original skew and surface heterogeneity. These visual patterns are consistent with the quantitative results: SAAM does not 
merely apply uniform smoothing, but uses cross-sectional attention to refine the representative forecast surface while maintaining a favorable balance between 
fit to market observations and static no-arbitrage residual control.

\subsection{Cross-Sectional Information Allocation in SAAM Refinement}
\label{subsec:attention_analysis}
We finally examine the cross-sectional attention weights learned by SAAM. This analysis is diagnostic rather than causal. It illustrates how the refinement operator allocates reference information across the implied volatility surface when adjusting a representative forecast. The main text focuses on ATM target coordinates, while the full attention matrices across additional moneyness regimes are reported in Appendix~\ref{app:attention_comprehensive}.

For a target coordinate $\mathbf{x}_i=(m_i,\tau_i)$, SAAM assigns attention weights over the reference grid $\mathcal{D}_0$. For a reference value field $v$, the resulting non-local aggregation can be written as
\begin{equation}
\mathcal{A}v(\mathbf{x}_i) = \sum_{j=1}^{N} \alpha(\mathbf{x}_i,\mathbf{x}_j)v(\mathbf{x}_j),
\label{eq:attention_integral_operator}
\end{equation}
where $v(\mathbf{x}_j)$ denotes the unrefined volatility value at reference coordinate $\mathbf{x}_j$, and $\alpha(\mathbf{x}_i,\mathbf{x}_j)$ is the normalized attention weight from $\mathbf{x}_i$ to $\mathbf{x}_j$. Since the heatmaps are scaled to their local maxima, they should be interpreted as relative routing patterns within each target configuration rather than as absolute magnitudes that can be compared directly across panels.

\begin{figure*}[!htbp]
\centering
\begin{minipage}{0.82\textwidth}
\centering
\includegraphics[width=\linewidth]{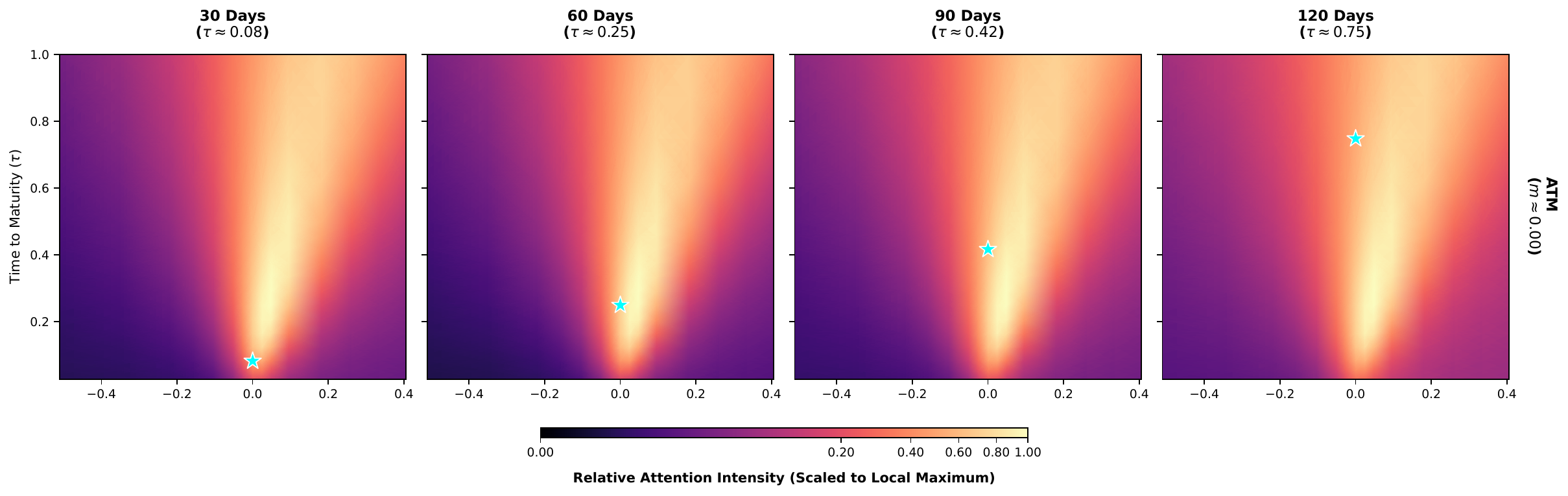}
\vspace{0.1cm}
\centerline{\small (a) Spatial Separation SAAM under the daily regime.}
\end{minipage}
\vspace{0.35cm}
\begin{minipage}{0.82\textwidth}
\centering
\includegraphics[width=\linewidth]{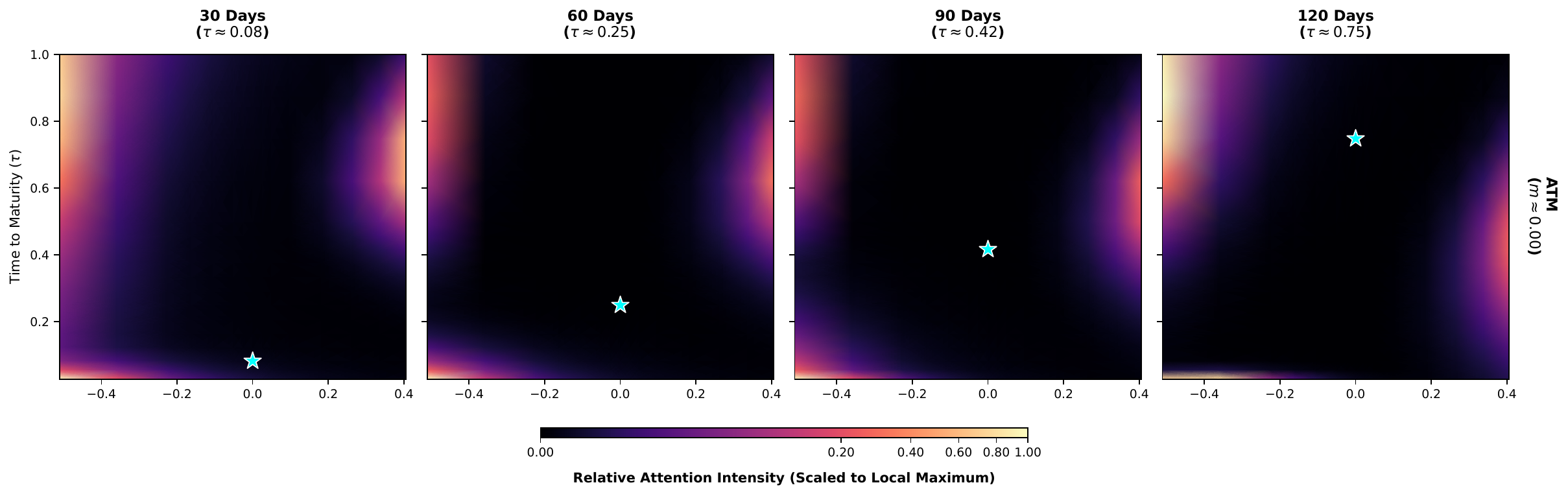}
\vspace{0.1cm}
\centerline{\small (b) Spatial Separation SAAM under the minute-level regime.}
\end{minipage}
\caption{Cross-attention weight heatmaps for Spatial Separation SAAM at ATM target coordinates. The star marks the target coordinate. Colors represent relative attention intensity scaled to the local maximum within each panel.}
\label{fig:separation_attention}
\end{figure*}

Figure~\ref{fig:separation_attention} reports the attention patterns of Spatial Separation SAAM. Since this variant uses only moneyness and maturity coordinates for attention routing, the heatmaps mainly reflect geometry-based information allocation. In the daily regime, attention is concentrated around the ATM region across maturities, with a particularly narrow focus at short maturities and a slightly wider range at longer maturities. This pattern is consistent with the role of maturity-wise information in the term structure of total implied variance and in calendar-spread diagnostics.

The minute-level pattern is more selective. The map becomes sparse and high-contrast: most interior regions receive little relative attention, while a small number of ATM, wing, and boundary coordinates are emphasized. This pattern is better interpreted as concentrated geometric routing rather than broad information dispersion. It suggests that coordinate-only routing becomes more selective at the minute frequency, relying on a limited set of reference locations when local high-frequency observations are less stable.

\begin{figure*}[!htbp]
\centering
\begin{minipage}{0.82\textwidth}
\centering
\includegraphics[width=\linewidth]{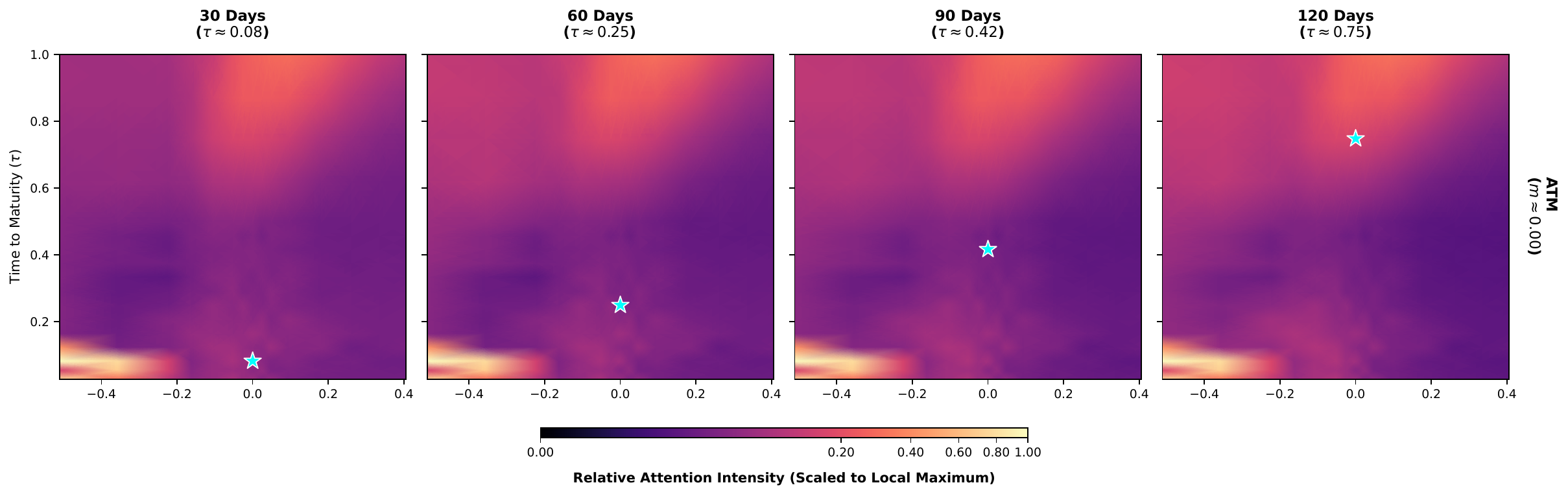}
\vspace{0.1cm}
\centerline{\small (a) Volatility Fusion SAAM under the daily regime.}
\end{minipage}
\vspace{0.35cm}
\begin{minipage}{0.82\textwidth}
\centering
\includegraphics[width=\linewidth]{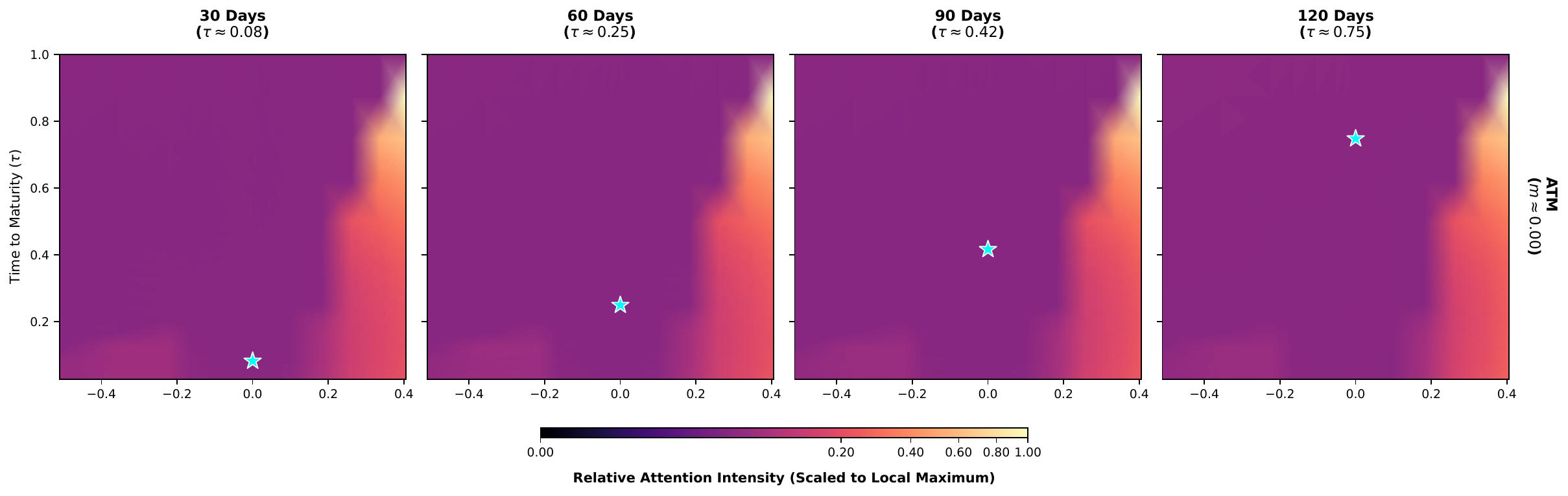}
\vspace{0.1cm}
\centerline{\small (b) Volatility Fusion SAAM under the minute-level regime.}
\end{minipage}
\caption{Cross-attention weight heatmaps for Volatility Fusion SAAM at ATM target coordinates. The star marks the target coordinate. Colors represent relative attention intensity scaled to the local maximum within each panel.}
\label{fig:fusion_attention}
\end{figure*}

Figure~\ref{fig:fusion_attention} shows the corresponding patterns for Volatility Fusion SAAM. This variant augments spatial coordinates with a volatility level proxy, so its routing can depend on both surface geometry and the prevailing volatility state. In the daily regime, the strongest attention is located in the short maturity left wing region, forming a distinctly left-skewed pattern. This region is economically informative because short-dated downside options are closely related to skew, tail risk demand, and stress-sensitive repricing. The pattern suggests that Volatility Fusion SAAM uses downside-sensitive regions as important references when coordinating the refinement of the ATM target.

At the minute-level, the high-intensity region expands toward both wings and boundary areas, producing a more balanced boundary emphasis than in the daily regime. Compared with Spatial Separation SAAM, the Fusion maps also retain a more visible low-intensity background across the surface. This difference indicates that the volatility level proxy does not simply select a few geometric anchors, but provides a broader state-dependent reference channel. Thus, Volatility Fusion SAAM combines boundary anchoring with a more diffuse cross-sectional context.

Taken together, Figures~\ref{fig:separation_attention} and~\ref{fig:fusion_attention} indicate that the two SAAM variants use cross-sectional information in different ways. Spatial Separation SAAM produces sharper geometry-based routing, especially at the minute-level, where attention is concentrated on selected reference locations. Volatility Fusion SAAM exhibits a more state-conditioned pattern, combining prominent wing or boundary anchors with a visible low-intensity background over the surface. These patterns are consistent with the quantitative results in Section~\ref{sec:refinement_results}: cross-sectional attention helps the refinement stage use non-local information from economically relevant regions of the IVS, improving the balance between fitting accuracy against market observations and finite-grid static no-arbitrage diagnostics.

\section{Conclusion}
\label{sec:conclusion}

This paper studies implied volatility surface forecasting as a risk surface modeling problem that requires both stochastic prediction and static no-arbitrage control. 
We propose a decoupled generative refinement framework. The diffusion stage learns the conditional distribution of future IVS states on a fixed grid of log-forward 
moneyness and maturity, while SAAM refines a representative forecast surface using market observations and static no-arbitrage residual diagnostics.

Empirical results on CSI 300 index options show that the conditional diffusion model improves  forecasting accuracy over  recurrent baselines 
and provides predictive intervals that capture heterogeneous uncertainty across moneyness, maturity, and sampling frequency. These advantages are particularly evident 
in the minute level  setting, where surface movements are more irregular.

The refinement results show that SAAM improves fitting accuracy at observed contract locations and reduces calendar spread and butterfly spread residual 
violations on the dense diagnostic grid. Compared with MLP refiners, the attention based refiners achieve a better balance between market observation fit 
and static no-arbitrage residual control. The attention diagnostics further suggest that Spatial Separation SAAM mainly captures geometry driven routing, 
whereas Volatility Fusion SAAM incorporates volatility level information into cross sectional information allocation.

Overall, the results support the use of a decoupled risk surface modeling pipeline for IVS forecasting. The diffusion stage provides distributional forecasts of 
future surface states, while the refinement stage improves the usability of the representative surface by balancing fit to market observations and static no-arbitrage 
residual control. This decomposition is relevant for practical option risk management, where probabilistic scenario information and stable representative pricing 
surfaces are both needed.

\begin{appendices}

    \section{Minute-Level Static No-Arbitrage Residual Violations in Raw Market Data}
    \label{app:minute level_violations}
    This appendix complements the daily evidence in Section~\ref{sec:exp_dataset} by reporting static no-arbitrage residual violations at the minute level. Compared with daily snapshots, minute-level observations are more affected by asynchronous trading, fragmented liquidity, and microstructure noise, which mainly appear as local convexity-related irregularities in the raw IVS states.
    
    Table~\ref{tab:minute_violations} reports summary statistics for minute-level residual violation rates in the raw CSI 300 data. These statistics quantify calendar spread, butterfly spread, and tail asymptotic irregularities observed prior to Stage II refinement.
    
    \begin{table}[htbp]
    \centering
    \caption{Summary of minute-level static no-arbitrage residual violation rates in the raw CSI 300 dataset.}
    \label{tab:minute_violations}
    \small
    \setlength{\tabcolsep}{5pt}
    \renewcommand{\arraystretch}{1.10}
    \begin{tabular}{lccccc}
    \hline
    Diagnostic & Mean & Median & Std & Min & Max \\
    \hline
    Calendar spread (C4) & $0.0006$ & $0.0000$ & $0.0016$ & $0.0000$ & $0.0134$ \\
    Butterfly (C5) & $0.1099$ & $0.1070$ & $0.0403$ & $0.0214$ & $0.2841$ \\
    Tail asymptotics (C6) & $0.0326$ & $0.0310$ & $0.0160$ & $0.0000$ & $0.1023$ \\
    \hline
    \end{tabular}
    \end{table}
    
    \begin{figure*}[htbp]
    \centering
    \includegraphics[width=0.95\textwidth]{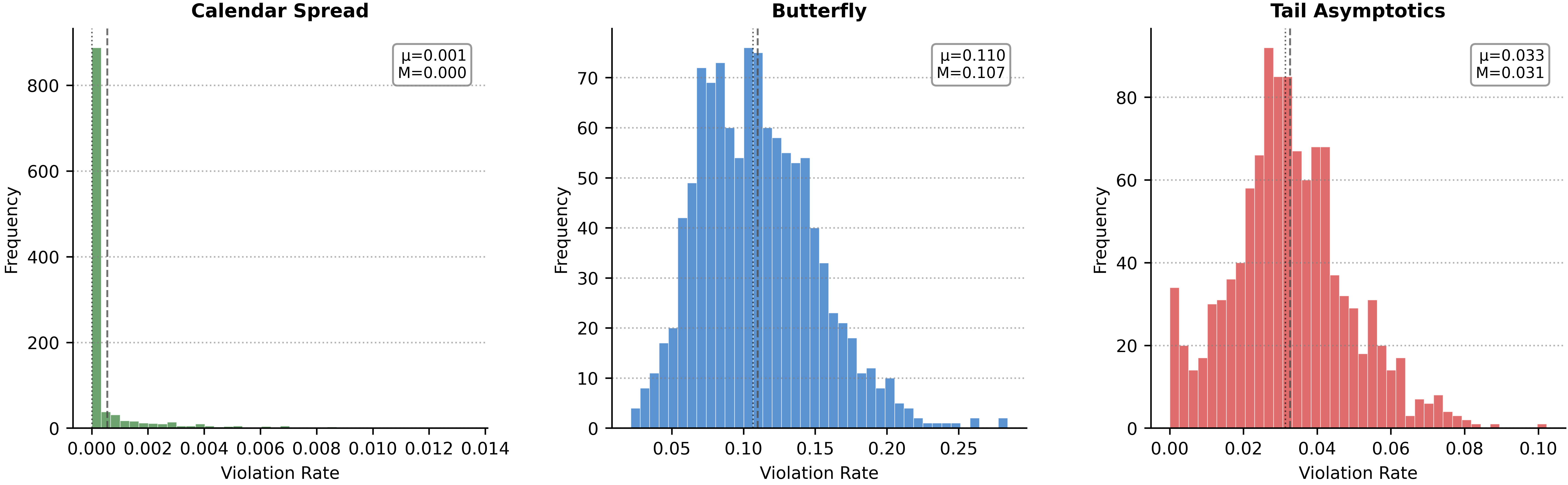}
    \caption{Empirical distributions of minute-level static no-arbitrage residual violation rates in the raw CSI 300 data.}
    \label{fig:app_violation_dist}
    \end{figure*}
    
    Table~\ref{tab:minute_violations} and Figure~\ref{fig:app_violation_dist} show that calendar spread violations remain concentrated near zero even at the minute level. In contrast, butterfly spread violations are much more pronounced, with a mean violation rate of about 11\% and a median of 10.7\%. Tail asymptotic violations are less pervasive but remain visible in the minute-level sample. This pattern suggests that minute-level market frictions mainly affect strike-wise convexity rather than term structure monotonicity.
    
    These diagnostics provide context for the minute-level Stage II refinement results in Section~\ref{sec:refinement_results}. They show that minute-level IVS forecasting should not be treated merely as scattered point fitting. A surface refinement mechanism is needed to reduce localized residual violations, especially those affecting strike-wise convexity and the implied risk-neutral density.

\section{Comprehensive Visualization of Cross Sectional Attention}
\label{app:attention_comprehensive}
Section~\ref{subsec:attention_analysis} examines attention routing at ATM target coordinates. This appendix extends the diagnostic view to $3 \times 4$ attention matrices across three moneyness regimes and four maturity horizons. The rows correspond to left wing, ATM, and right wing target coordinates, while the columns correspond to increasing maturities. These figures show how the routing patterns vary when the target coordinate moves across the implied volatility surface.

\subsection{Spatial Separation SAAM}
\begin{figure}[p]
\centering
\begin{minipage}{0.98\textwidth}
\centering
\includegraphics[width=\linewidth,height=0.38\textheight,keepaspectratio]{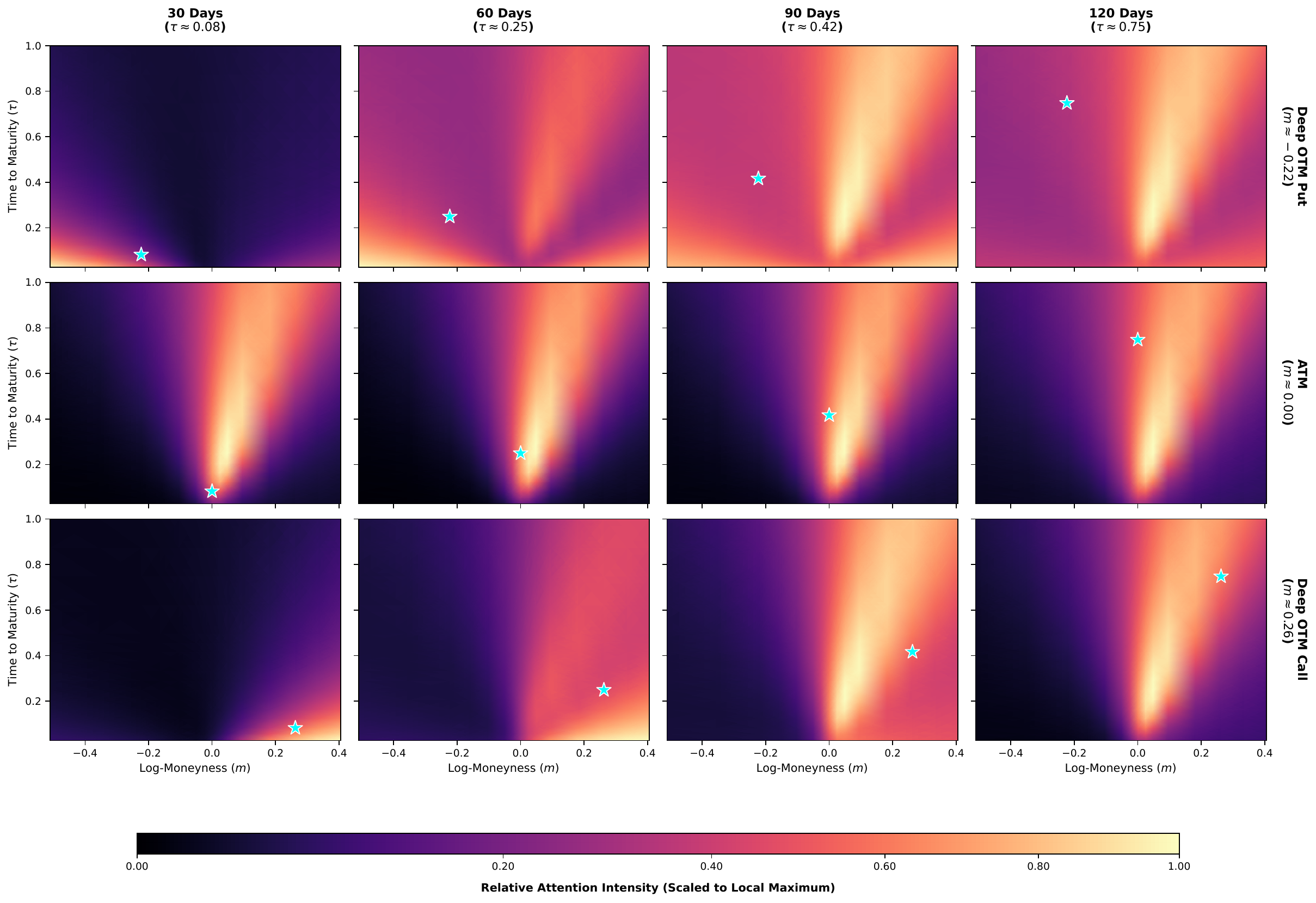}
\vspace{0.1cm}
\centerline{\small (a) Daily regime.}
\end{minipage}
\vspace{0.35cm}
\begin{minipage}{0.98\textwidth}
\centering
\includegraphics[width=\linewidth,height=0.38\textheight,keepaspectratio]{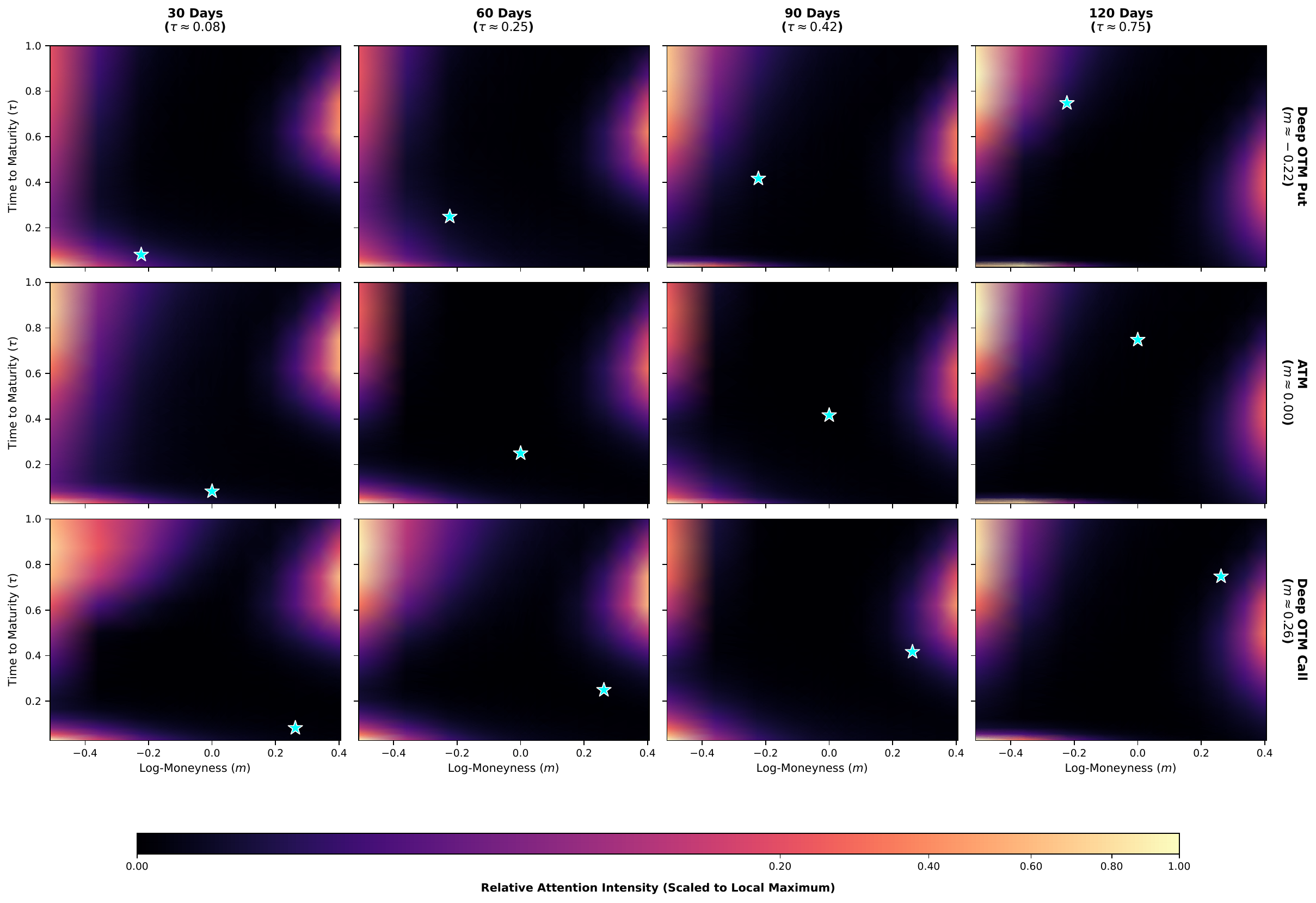}
\vspace{0.1cm}
\centerline{\small (b) Minute Level regime.}
\end{minipage}
\caption{Comprehensive $3 \times 4$ attention heatmaps for Spatial Separation SAAM. Rows correspond to left wing, ATM, and right wing target coordinates; columns correspond to increasing maturities. The star marks the target coordinate in each panel.}
\label{fig:app_separation_attention}
\end{figure}

Figure~\ref{fig:app_separation_attention} reports the full attention matrices for Spatial Separation SAAM. Since this variant uses only moneyness and maturity coordinates for attention routing, the heatmaps mainly reflect geometry based information allocation.

In the daily regime, the routing pattern varies with the target location. For ATM targets, attention is concentrated around maturity wise reference regions, which is consistent with the role of the term structure of total implied variance in calendar spread diagnostics. For wing targets, the reference pattern shifts with the target moneyness and maturity. This behavior is consistent with the design of Spatial Separation SAAM, where routing is governed mainly by coordinate geometry rather than by a common volatility level signal.

At the minute level frequency, the pattern becomes sharper. Across different target moneyness levels, most interior regions receive low relative attention, while selected boundary or wing regions become more prominent. This complements the ATM based observation in Section~\ref{subsec:attention_analysis}. For Spatial Separation SAAM, minute level routing is better described as selective boundary anchoring than as diffuse global information aggregation.

\subsection{Volatility Fusion SAAM}

\begin{figure}[p]
\centering
\begin{minipage}{0.98\textwidth}
\centering
\includegraphics[width=\linewidth,height=0.38\textheight,keepaspectratio]{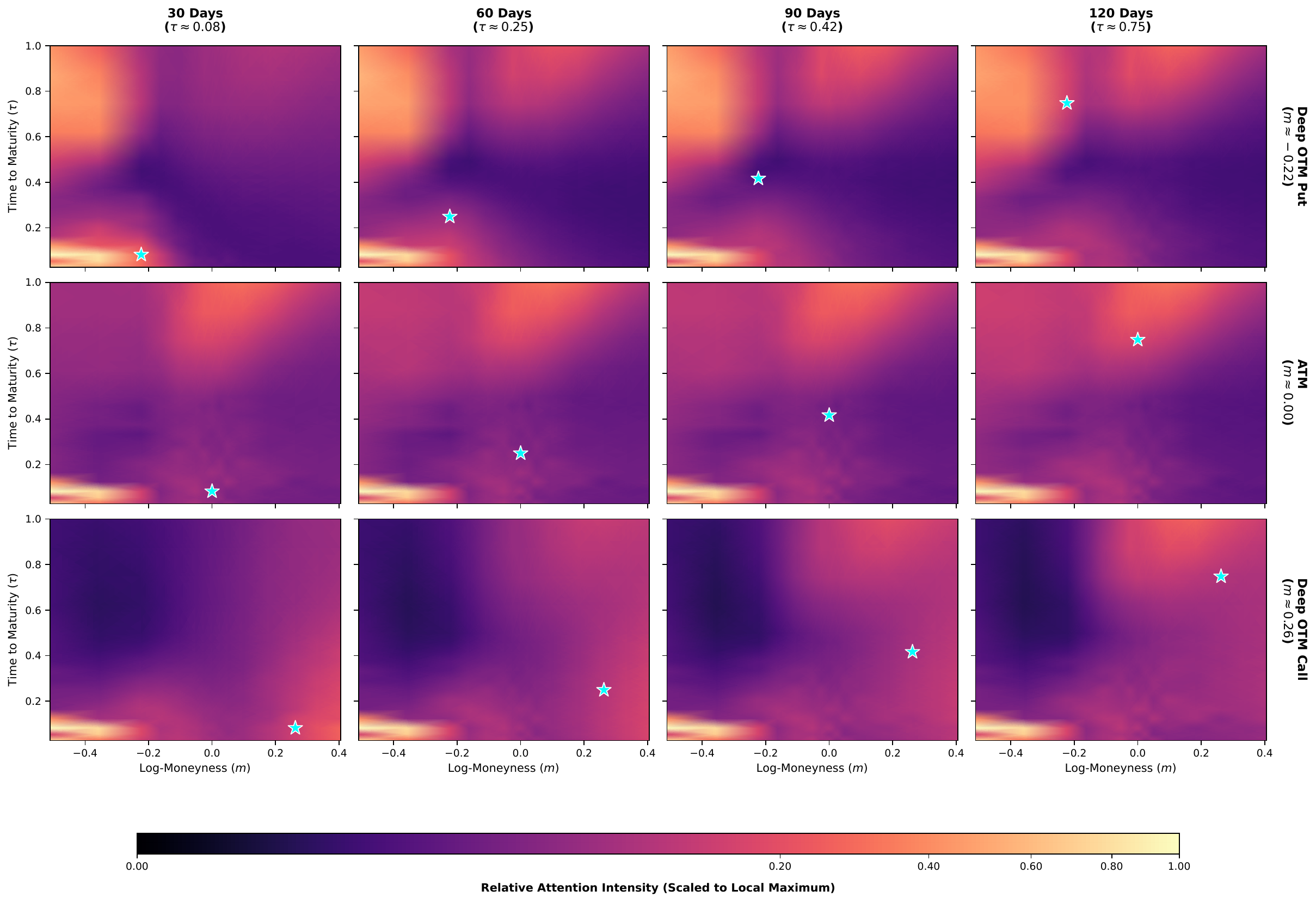}
\vspace{0.1cm}
\centerline{\small (a) Daily regime.}
\end{minipage}
\vspace{0.35cm}
\begin{minipage}{0.98\textwidth}
\centering
\includegraphics[width=\linewidth,height=0.38\textheight,keepaspectratio]{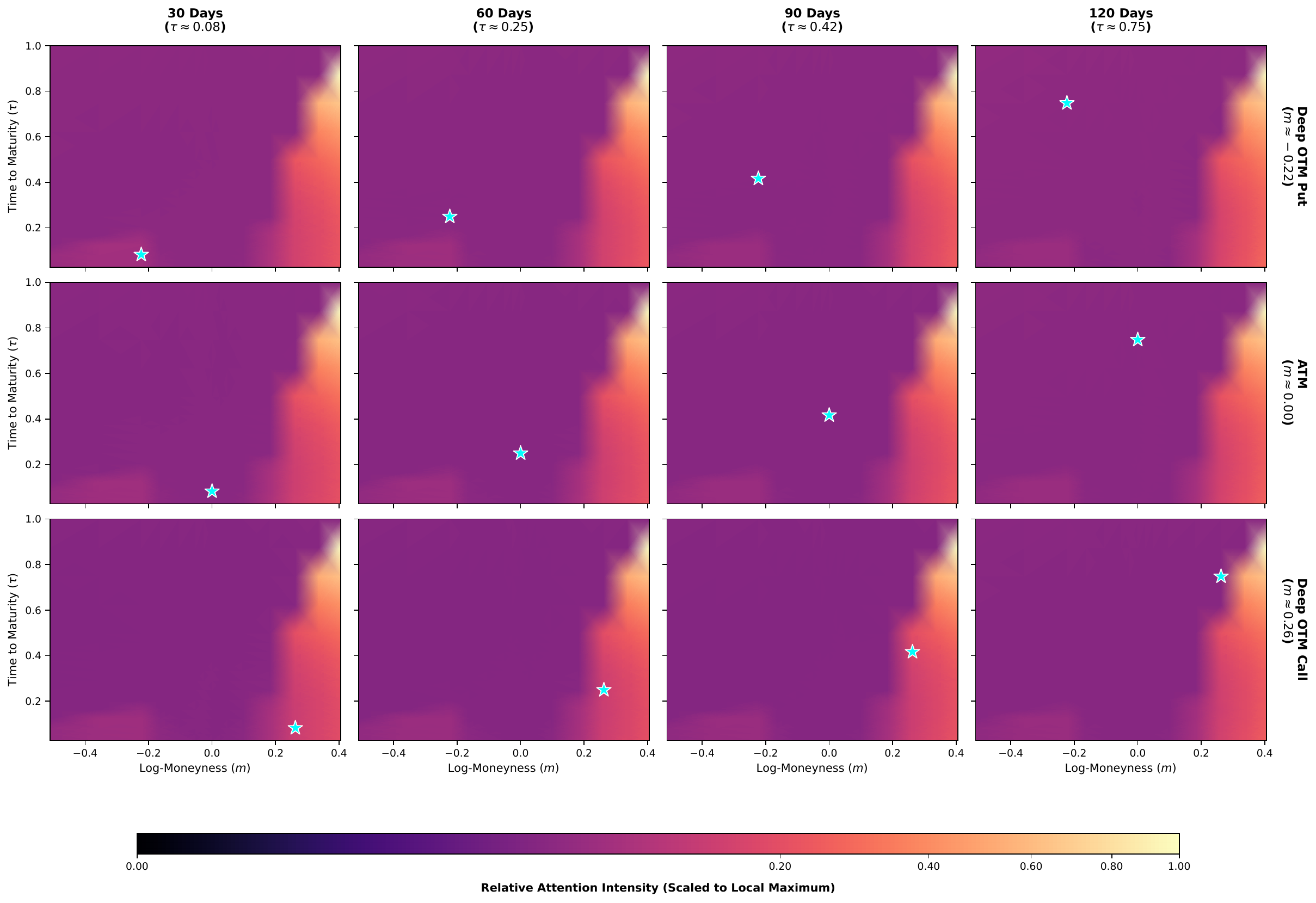}
\vspace{0.1cm}
\centerline{\small (b) Minute Level regime.}
\end{minipage}
\caption{Comprehensive $3 \times 4$ attention heatmaps for Volatility Fusion SAAM. Rows correspond to left wing, ATM, and right wing target coordinates; columns correspond to increasing maturities. The star marks the target coordinate in each panel.}
\label{fig:app_fusion_attention}
\end{figure}
Figure~\ref{fig:app_fusion_attention} reports the corresponding full attention matrices for Volatility Fusion SAAM. This variant augments the spatial coordinates with a volatility level proxy, allowing the routing weights to depend on both target geometry and the volatility state of the representative surface.

In the daily regime, the Fusion maps vary with the target coordinate but retain visible attention to wing and boundary regions. This complements the main text ATM result, where the short maturity left wing is an important reference region. The pattern is consistent with the design of the Fusion variant, in which the volatility level proxy introduces a surface level conditioning signal in addition to coordinate geometry.

At the minute level frequency, the Fusion maps also place strong attention on boundary or wing regions. Compared with Spatial Separation SAAM, however, they retain a more visible low intensity background across the grid. This suggests that the volatility level proxy provides a broader state conditioned reference channel, while still allowing the operator to use specific boundary regions as anchors when refining high frequency surfaces.

Overall, the comprehensive matrices complement the ATM based analysis in Section~\ref{subsec:attention_analysis}. They show that the distinction between the two SAAM variants remains visible across moneyness and maturity levels. Spatial Separation SAAM mainly reflects geometry based routing, with target dependent reference patterns and selective boundary anchoring at the minute level frequency. Volatility Fusion SAAM combines wing or boundary anchors with a more diffuse volatility conditioned background. These patterns support the interpretation of SAAM as a data adaptive cross sectional refinement rule rather than a fixed local smoothing filter.

\section{Penalty Weight Sensitivity}
\label{app:penalty_weight_selection}

\begin{figure}[htbp]
\centering
\includegraphics[width=0.95\textwidth]{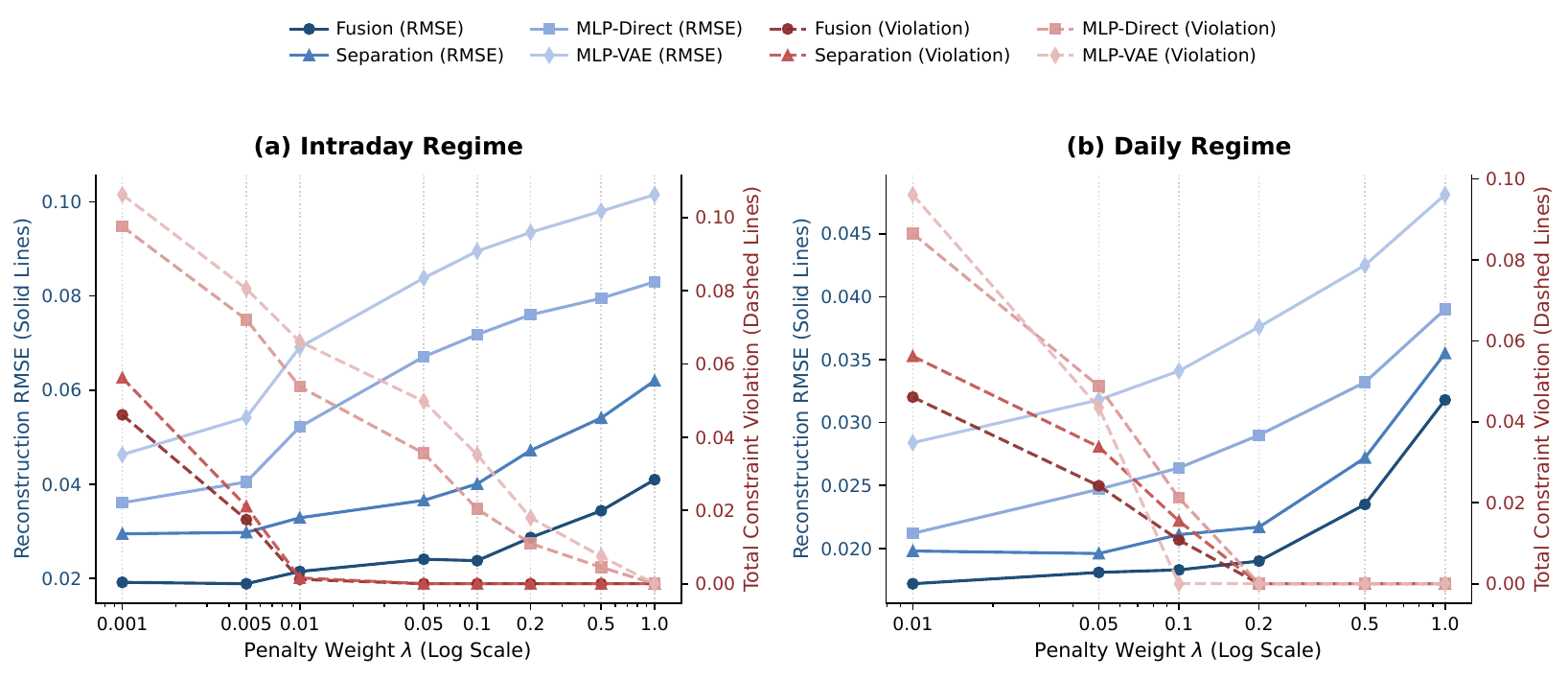}
\caption{Representative sensitivity of reconstruction RMSE and total residual violations to the penalty weight $\lambda$. Solid lines report RMSE, while dashed lines report total residual violations.}
\label{fig:app_lambda_tradeoff}
\end{figure}

To contextualize the hyperparameter configurations used in Table~\ref{tab:SAAM_vs_DNN_Comprehensive}, Figure~\ref{fig:app_lambda_tradeoff} provides a representative illustration of the tradeoff between reconstruction accuracy and static no-arbitrage residual control during model development. 

The selected operating points reported in Table~\ref{tab:SAAM_vs_DNN_Comprehensive} correspond to the final configurations adopted for formal evaluation. Table~\ref{tab:lambda_selection} summarizes the penalty weights used in these configurations.

\begin{table}[htbp]
\centering
\caption{Penalty weights used in the Stage II refinement experiments.}
\label{tab:lambda_selection}
\small
\setlength{\tabcolsep}{6pt}
\renewcommand{\arraystretch}{1.10}
\begin{tabular}{lcc}
\hline
Method & Minute-level & Daily \\
\hline
Volatility Fusion SAAM & $0.01$ & $0.20$ \\
Spatial Separation SAAM & $0.01$ & $0.20$ \\
MLP Direct & $1.00$ & $0.20$ \\
MLP VAE & $1.00$ & $0.10$ \\
\hline
\end{tabular}
\end{table}

As illustrated in the minute-level setting in Figure~\ref{fig:app_lambda_tradeoff}(a), the deterministic baselines require substantially larger penalty weights to reduce static no-arbitrage residual violations. At lower penalty magnitudes, residual violations remain comparatively large, indicating weak residual control under severe microstructure noise. For the MLP refiners, the selected penalty weight is $\lambda=1.0$, which reduces residual violations but also increases reconstruction RMSE, suggesting that pointwise refinement becomes less stable under strong residual penalization.

The SAAM variants exhibit a more favorable tradeoff at the minute-level. Smaller penalty weights yield slightly lower reconstruction errors but leave nonnegligible residual violations. Increasing the penalty to the selected operating point, $\lambda=0.01$, substantially reduces residual violations while preserving strong reconstruction accuracy. This pattern is consistent with the role of cross-sectional attention in incorporating surface-level context during refinement.

A similar but milder tradeoff is observed in the daily setting in Figure~\ref{fig:app_lambda_tradeoff}(b). In this smoother regime, $\lambda=0.20$ is sufficient for the SAAM 
variants and the MLP Direct baseline, while MLP VAE uses $\lambda=0.10$. The key difference lies in reconstruction accuracy. Although the MLP baselines can reduce residual 
violations at these selected penalty weights, they incur larger RMSE values. By contrast, the SAAM variants maintain lower RMSE for example, $0.0190$ for Volatility Fusion 
SAAM, while keeping residual violations small.

Overall, Figure~\ref{fig:app_lambda_tradeoff} serves as a representative sensitivity analysis rather than an exhaustive hyperparameter study. It shows that the penalty weight directly controls the balance between reconstruction accuracy and finite-grid static no-arbitrage residual control. The comparison also supports the use of attention-based refiners, which achieve a more favorable balance in both daily and minute-level settings.

\end{appendices}

\bmhead{Statements and Declarations}

\begin{itemize}
    \item \textbf{Funding:} This work was supported by the National Key R\&D Program of China (No. 2023YFA1008701) and the Key Project of the National Natural Science Foundation of China (No. 12431017).
    
    \item \textbf{Competing interests:} The authors declare that they have no competing interests.
    
    \item \textbf{Data availability:} The CSI 300 index option data used in this study were obtained from RiceQuant (\url{https://www.ricequant.com/}). The raw data are subject to the licensing terms of the data provider and cannot be publicly redistributed by the authors. Researchers may obtain access to the same data source through RiceQuant, subject to the provider's subscription and licensing conditions.
    
    \item \textbf{Code availability:} The code used for model training and empirical analysis is available from the corresponding author upon reasonable request.
    
    \item \textbf{Author contributions:} \textbf{Lifeng Hao:} Conceptualization, methodology, software, validation, formal analysis, investigation, data curation, writing original draft, and visualization. \textbf{Shaolin Ji:} Conceptualization, methodology, resources, writing review and editing, supervision, project administration, and funding acquisition. Both authors contributed equally to this work, read and approved the final manuscript.
\end{itemize}

\bibliography{sn-bibliography}

\end{document}